\documentclass[12pt]{article}

\usepackage{epsfig} 

\usepackage{authblk}
\usepackage{cmll}
\usepackage{dsfont}
\usepackage{array}
\usepackage{cite}
\usepackage{amsfonts, amssymb, amsthm}
\usepackage{amsmath, accents, bbold}

\newcommand{\be}{\begin{equation}}
\newcommand{\ee}{\end{equation}}
\newcommand{\bea}{\begin{eqnarray}}
\newcommand{\eea}{\end{eqnarray}}

\begin{document}

\title{\bf The Weyl -- Wigner -- Moyal Formalism on a Discrete Phase Space. II. The Photon Wigner Function}

\author[1]{ Maciej Przanowski \footnote{Professor Emeritus} \thanks{e.mail: maciej.przanowski@p.lodz.pl}} 
\author[1] {Jaromir Tosiek  \thanks{e.mail: jaromir.tosiek@p.lodz.pl}} 
\author[2] { Francisco J. Turrubiates \thanks{e.mail: fturrub@esfm.ipn.mx} }
  
 \affil[1]{{\footnotesize{\em
Institute of Physics, \L\'{o}d\'{z} University of Technology,  90-924 \L\'{o}d\'{z},   Poland,}}}
\affil[2]{\footnotesize{\em Departamento de F\'{\i}sica, Escuela Superior de F\'{\i}sica
y Matem\'aticas,  Instituto Polit\'{e}cnico Nacional, Unidad Adolfo
L\'opez Mateos, Edificio 9,  07738  Ciudad de M\'exico , M\'exico}}
 
 \date{\today}
 
 \maketitle

\begin{abstract}

Classical model of light  in helicity formalism is presented. Then quantum point of view at photons -- construction and interpretation of photon wave function is proposed.   
Quantum mechanics of photon is investigated. The Bia\l ynicki -- Birula scalar product $\langle {\bf \Psi}_{1}|{\bf \Psi}_{2}\rangle_{BB}$ and the generalized Hermitian conjugation  $\widehat{\gamma}^{+\hspace{-0.45em}+}$ of linear operator $\widehat{\gamma}$ are discussed. Quantum description of light on a phase space
 is developed. A photon Wigner function is built. 
\end{abstract}

\section{Introduction}
This paper is the second part of our work devoted to the Weyl -- Wigner -- Moyal formalism for particles with discrete internal degrees of freedom. In the first part \cite{Przanowski2019} we  developed a general approach and then we applied it to a nonrelativistic particle of spin $\frac{1}{2}$ in a homogeneous magnetic field and to  analysis of magnetic resonance for a spin $\frac{1}{2}$ uncharged nonrelativistic particle endowed with a magnetic moment. 

In the present article we study the Weyl -- Wigner -- Moyal formalism for a photon and, in particular, we find the photon Wigner function which arises in a natural way within this construction.

Quantum mechanics of photon which we adopt in this paper, was developed by Iwo Bia\l ynicki -- Birula \cite{Bialynicki1994, Bialynicki1996, Bialynicki2006} and John E. Sipe \cite{Sipe1995} (see also \cite{Bialynicki2017, Chandrasekar2012}). Given this quantum mechanics a natural question that can be asked is how one can construct a phase space picture of such quantum mechanics  in the Weyl -- Wigner -- Moyal sense and, in particular, how to define a respective Wigner function. The photon Wigner function within quantum mechanics was proposed by I. Bia\l ynicki -- Birula in \cite{Bialynicki1994, Bialynicki1996}. However, it has been constructed by analogy to the usual nonrelativistic case and not as a consequence of any Weyl -- Wigner -- Moyal formalism for the photon. The same comment concerns also the photon Wigner function defined in \cite{Bialynicki2014} with the use of field operators corresponding to the Riemann -- Silberstein vector and its conjugate.

In the present paper we adopt a general theory of the Weyl -- Wigner -- Moyal formalism of quantum particles with internal degrees of freedom \cite{Przanowski2019, Przanowski2017} (see also references therein) to the photon. Thus one deals with a massless relativistic particle of spin $1$ but with an additional constraint saying that the helicity of photon can take only two values: $+1$ or $-1.$ So the respective Hilbert space is ${\mathcal H}= L^2({\mathbb R}^3) \otimes {\mathbb C}^3$ with the subsidiary condition.

The paper is organised as follows. In Section \ref{sec2} we recall some facts from classical Maxwell electrodynamics in vacuum which are pertinent for further considerations. We use the spinorial and helicity formalisms. The Riemann -- Silberstein vector is widely employed as the main object appearing in the Maxwell equations. This vector plays a crucial role in formulating quantum mechanics of photon and in defining a photon wave function. 

In Sec. \ref{sec3} the photon wave function is defined and its probability interpretation is given. We mainly follow the works \cite{Bialynicki1994, Bialynicki1996, Sipe1995}. We consider also a unitary transformation leading to the spinorial representation of the photon wave function.

Section \ref{sec4} is devoted to further study of structure of photon quantum mechanics. We investigate the Bia\l ynicki -- Birula scalar product and a connection between observables and generalised Hermitian operators. Then we define a density operator for the photon and find its properties. 

The Weyl -- Wigner -- Moyal formalism for the photon is developed in Sec. \ref{sec5}. We introduce a photon Wigner function and study its features.

Some specific examples of photon Wigner functions are presented in Sec. \ref{sec6}. Final remarks are presented in Sec. \ref{sec7}. The paper is ended with Appendix \ref{appendixA} containing explicit formulae for $*$  -- products on phase space ${\mathbb R}^6$ and on the grid $\Gamma^3.$


\section{The classical Maxwell electromagnetic field. The spinorial and helicity formalisms}
\label{sec2}
In this section we describe the classical Maxwell electromagnetic field in vacuum within the spinorial and helicity formalisms \cite{Bialynicki1994, Bialynicki1996, Plebanski1974, Plebanski1980, Corson1954, Penrose1984}. We deal with the Minkowski spacetime of signature $(+,+,+,-).$ In the Lorentz coordinate system $(x^1,x^2,x^3,x^4=ct)$ the line element reads
\begin{eqnarray}\label{21}
  ds^{2} &=& \eta_{\mu\nu}dx^{\mu}dx^{\nu} 
   = (dx^{1})^{2}+(dx^{2})^{2}+(dx^{3})^{2}-(dx^{4})^{2} \nonumber \\
     &=& d\vec{x}^{\;2} - c^{2}dt^{2},
\end{eqnarray}
where
\begin{equation*}
  \left(\eta_{\mu\nu}\right)=\left(
                               \begin{array}{cccc}
                                 1 & 0 & 0 & 0 \\
                                  0 & 1 & 0 & 0 \\
                                  0 & 0 & 1 & 0 \\
                                  0 & 0 & 0 & -1 \\
                               \end{array}
                             \right)
                             =\left(\eta^{\mu\nu}\right), \,\,\, \mu,\nu=1,2,3,4.
\end{equation*}
Define a spinorial $1$ -- form
\begin{equation}\label{22}
  g^{A\accentset{\mbox{\huge .} }{B}}:=g_{\mu}^{\enspace A  \accentset{\mbox{\huge .}} {B}}dx^{\mu}, \quad A=1,2, \enspace \accentset{\mbox{\huge .}}{B}= \accentset{\mbox{\huge .}}{1}, \accentset{\mbox{\huge .}}{2}
\end{equation}
with  matrices $\left( g_{\mu}^{\enspace A  \accentset{\mbox{\huge .}}{B}} \right)$ given by
\begin{equation}\label{23}
  \left( g_{\mu}^{\enspace A  \accentset{\mbox{\huge .}}{B}} \right):=\left\{ \left(
                                         \begin{array}{cc}
                                           0 & 1 \\
                                           1 & 0 \\
                                         \end{array}
                                       \right), \left(
                                                  \begin{array}{cc}
                                                    0 & -i \\
                                                    i & 0 \\
                                                  \end{array}
                                                \right), \left(
                                                           \begin{array}{cc}
                                                             1 & 0 \\
                                                             0 & -1 \\
                                                           \end{array}
                                                         \right),
                                                          \left(
                                \begin{array}{cc}
                                  -1 & 0 \\
                                  0 & -1 \\
                                \end{array}
                              \right)
   \right\}.
\end{equation}
Hence by (\ref{21})
\begin{equation}\label{24}
  \left( g^{\mu  \,A  \accentset{\mbox{\huge .}}{B}} \right):=\left\{ \left(
                                         \begin{array}{cc}
                                           0 & 1 \\
                                           1 & 0 \\
                                         \end{array}
                                       \right), \left(
                                                  \begin{array}{cc}
                                                    0 & -i \\
                                                    i & 0 \\
                                                  \end{array}
                                                \right), \left(
                                                           \begin{array}{cc}
                                                             1 & 0 \\
                                                             0 & -1 \\
                                                           \end{array}
                                                         \right),
                                                          \left(
                                \begin{array}{cc}
                                  1 & 0 \\
                                  0 & 1 \\
                                \end{array}
                              \right)
   \right\}.
\end{equation}
Then one introduces symmetric spinorial $2$ -- forms 
\begin{subequations}\label{25}
\begin{align}
   S^{AB}&:= \frac{1}{2}\epsilon_{\accentset{\mbox{\huge .}}{C}\accentset{\mbox{\huge .}}{D}} \,g^{A\accentset{\mbox{\huge .}}{C}}\wedge g^{B\accentset{\mbox{\huge .}}{D}}  =  S^{BA},
   \label{25a} \\
   S^{\accentset{\mbox{\huge .}}{A}\accentset{\mbox{\huge .}}{B}}&:= \frac{1}{2}\epsilon_{CD} g^{C\accentset{\mbox{\huge .}}{A}}\wedge g^{D\accentset{\mbox{\huge .}}{B}}=(S^{AB})^{\ast}= 
   S^{\accentset{\mbox{\huge .}}{B}\accentset{\mbox{\huge .}}{A}},  \label{25b} \\
   A,B, &C,D=1,2; \enspace \accentset{\mbox{\huge .}}{A},\accentset{\mbox{\huge .}}{B},\accentset{\mbox{\huge .}}{C},\accentset{\mbox{\huge .}}{D}=\accentset{\mbox{\huge .}}{1},\accentset{\mbox{\huge .}}{2} \nonumber
\end{align}
\end{subequations}
where $\epsilon_{CD}=\epsilon^{CD}$ and $\epsilon_{\accentset{\mbox{\huge .}}{C}\accentset{\mbox{\huge .}}{D}}=\epsilon^{\accentset{\mbox{\huge .}}{C}\accentset{\mbox{\huge .}}{D}}$ are antisymmetric spinors (the Levi-Civita symbols)
\begin{equation}\label{26}
  (\epsilon_{CD})=(\epsilon^{CD})=\left(
                                    \begin{array}{cc}
                                      0 & 1 \\
                                      -1 & 0 \\
                                    \end{array}
                                  \right) = (\epsilon_{\accentset{\mbox{\huge .}}{C}\accentset{\mbox{\huge .}}{D}})=(\epsilon^{\accentset{\mbox{\huge .}}{C}\accentset{\mbox{\huge .}}{D}})
\end{equation}
and the asterix ``$\ast$'' stands for the complex conjugation.

[An explicit expression for  $2$ -- form $S^{AB}$ is
\[
(S^{AB})= \frac{1}{2} \big(S_{\mu \nu}^{\enspace  AB}\big) dx^{\mu} \wedge dx^{\nu}=
\left( \begin{array}{cc}
                                      0 & -i \\
                                      -i & 0 \\
                                    \end{array}
                                  \right) dx^1 \wedge dx^2 + \left( \begin{array}{cc}
                                      -1 & 0 \\
                                      0 & -1 \\
                                    \end{array}
                                  \right) dx^1 \wedge dx^3 
\]
\[
+\left( \begin{array}{cc}
                                      1 & 0 \\
                                      0 & -1 \\
                                    \end{array}
                                  \right) dx^1 \wedge dx^4
+ \left( \begin{array}{cc}
                                      i & 0 \\
                                      0 & -i \\
                                    \end{array}
                                  \right) dx^2 \wedge dx^3 
 \]
 \[                                
                                  + \left( \begin{array}{cc}
                                      -i & 0 \\
                                      0 & -i \\
                                    \end{array}
                                  \right) dx^2 \wedge dx^4 +\left( \begin{array}{cc}
                                      0 & -1 \\
                                      -1 & 0 \\
                                    \end{array}
                                  \right) dx^3 \wedge dx^4.
\]
 $S^{\accentset{\mbox{\huge .}}{A}\accentset{\mbox{\huge .}}{B}}$ is automatically generated from $S^{AB}$ via (\ref{25b}).]

Spinorial indices are manipulated with the use of $\epsilon_{\accentset{\mbox{\huge .}}{A}\accentset{\mbox{\huge .}}{B}}$ and  $\epsilon^{\accentset{\mbox{\huge .}}{A}\accentset{\mbox{\huge .}}{B}}$ as follows
\begin{equation}\label{27}
  \chi_{\accentset{\mbox{\huge .}}{A}}=\epsilon_{\accentset{\mbox{\huge .}}{A}\accentset{\mbox{\huge .}}{B}}\,\chi^{\accentset{\mbox{\huge .}}{B}} \quad, \quad 
  \chi^{\accentset{\mbox{\huge .}}{A}}=\epsilon^{\accentset{\mbox{\huge .}}{B}\accentset{\mbox{\huge .}}{A}}\,\chi_{\accentset{\mbox{\huge .}}{B}}
\end{equation}
and analogously one defines the lowering and rising indices for undotted spinorial indices with the use of $\epsilon_{AB}$ and $\epsilon^{AB}.$

Equations  (\ref{25}) are equivalent to the formula 
\begin{equation}\label{28}
  g^{A \accentset{\mbox{\huge .}}{B}}\wedge g^{C \accentset{\mbox{\huge .}}{D}}= \epsilon^{\accentset{\mbox{\huge .}}{B}\accentset{\mbox{\huge .}}{D}}S^{AC}+\epsilon^{AC}S^{\accentset{\mbox{\huge .}}{B}\accentset{\mbox{\huge .}}{D}}.
\end{equation}
We adapt the definition of  {\it the Hodge star operation} (the Hodge -- $\ast$) to 2 -- forms as
\begin{equation}\label{29}
  \omega=\frac{1}{2}\omega_{\mu\nu}dx^{\mu}\wedge dx^{\nu} \quad \Longrightarrow \quad \ast \omega:=\frac{1}{2}\left(- \frac{i}{2}\epsilon_{\rho\sigma\mu\nu}\omega^{\mu\nu}dx^{\rho}\wedge dx^{\sigma}\right)
\end{equation}
with $\epsilon_{\rho\sigma\mu\nu}$ being the Levi -- Civita totally antisymmetric symbol  $\epsilon_{1234}=1$. Indices of $\omega_{\mu\nu}$ have been raised according to the rule $\omega^{\mu\nu}= \eta^{\mu \alpha} \eta^{\nu \beta} \omega_{\alpha \beta}.$

One quickly finds that
\be
\label{210}
\ast  \ast \omega= \omega
\ee
and
\be
\label{211}
\ast S^{AB}= S^{AB} \;\;\;, \;\;\; \ast   S^{\accentset{\mbox{\huge .}}{A}\accentset{\mbox{\huge .}}{B}}= -  S^{\accentset{\mbox{\huge .}}{A}\accentset{\mbox{\huge .}}{B}}.
\ee
Therefore $S^{AB}$ are  self -- dual $2$ -- forms and $ S^{\accentset{\mbox{\huge .}}{A}\accentset{\mbox{\huge .}}{B}}$ are  anti -- self -- dual $2$ -- forms. In fact the system of three 
$2$ -- forms $ \left\{ S^{11}, S^{12}=S^{21}, S^{22} \right\}$
 constitutes a basis of vector space of self -- dual $2$ -- forms and the system 
 $ \left\{
 S^{\accentset{\mbox{\huge .}}{1}\accentset{\mbox{\huge .}}{1}},
 S^{\accentset{\mbox{\huge .}}{1}\accentset{\mbox{\huge .}}{2}}=
 S^{\accentset{\mbox{\huge .}}{2}\accentset{\mbox{\huge .}}{1}},
 S^{\accentset{\mbox{\huge .}}{2}\accentset{\mbox{\huge .}}{2}} \right\}
 $ is a basis of   linear space of anti -- self -- dual $2$ -- forms.
  
  Let
\begin{equation}\label{212}
 \mathcal{A}=\mathcal{A}_{\mu}dx^{\mu} \quad, \quad \mathcal{A}_{\mu}=(\vec{A},-\varphi)
\end{equation}
be an \textit{electromagnetic potential $1$ -- form} and
\begin{align}\label{213}
  \mathcal{F}&= d\mathcal{A}=\frac{1}{2}\mathcal{F}_{\mu\nu}dx^{\mu}\wedge dx^{\nu}, \nonumber \\
  \mathcal{F}_{\mu\nu}&=\partial_{\mu}\mathcal{A}_{\nu}-\partial_{\nu}\mathcal{A}_{\mu}; \,\,\, \partial_{\mu}\equiv\frac{\partial}{\partial x^{\mu}}
\end{align}
the \textit{electromagnetic field $2$ -- form}.  Using  definition (\ref{212}) of $\mathcal{A}_{\mu}$ one easily gets
\begin{equation}\label{215}
  (\mathcal{F}_{\mu\nu})=\left(
                           \begin{array}{cccc}
                             0 & \mathcal{B}_{3} & -\mathcal{B}_{2} & \mathcal{E}_{1} \\
                             -\mathcal{B}_{3} & 0 & \mathcal{B}_{1} & \mathcal{E}_{2} \\
                             \mathcal{B}_{2} & -\mathcal{B}_{1} & 0 & \mathcal{E}_{3} \\
                             -\mathcal{E}_{1} & -\mathcal{E}_{2} & -\mathcal{E}_{3} & 0 \\
                           \end{array}
                         \right),
\end{equation}
where $\vec{\mathcal{E}}=$($\mathcal{E}_{1}, \mathcal{E}_{2}, \mathcal{E}_{3}$) and $\vec{\mathcal{B}}=$($\mathcal{B}_{1}, \mathcal{B}_{2}, \mathcal{B}_{3}$) are vectors of  electric and magnetic fields respectively.

We express  $2$ -- form $\mathcal{F}$ in terms of  $S^{AB}$  and $ S^{\accentset{\mbox{\huge .}}{A}\accentset{\mbox{\huge .}}{B}}$ as
\begin{align}\label{215}
  \mathcal{F}&=\frac{1}{2}\left(f_{AB}S^{AB} + f_{\accentset{\mbox{\huge .}}{A}\accentset{\mbox{\huge .}}{B}}S^{\accentset{\mbox{\huge .}}{A}\accentset{\mbox{\huge .}}{B}}\right)\\
  &f_{AB}=f_{BA}, \quad f_{\accentset{\mbox{\huge .}}{A}\accentset{\mbox{\huge .}}{B}}=(f_{AB})^{\ast}=f_{\accentset{\mbox{\huge .}}{B}\accentset{\mbox{\huge .}}{A}} \nonumber
\end{align}
[Observe that formula (\ref{215}) differs from the respective one in Pleba\'{n}ski's notes \cite{Plebanski1974, Plebanski1980} for the factor $\frac{1}{2},$ which we use here for further convenience.]

In order to calculate  coefficients $f_{AB}$ and $f_{\accentset{\mbox{\huge .}}{A}\accentset{\mbox{\huge .}}{B}}$ we introduce spin -- tensors  $S_{\enspace \enspace \mu \nu}^{  AB} $ and $S^{\accentset{\mbox{\huge .}}{A}\accentset{\mbox{\huge .}}{B}}_{\enspace \enspace \mu \nu}$ by
\be
\label{215ad}
S^{AB}=\frac{1}{2}S_{\enspace \enspace \mu \nu}^{  AB} dx^{\mu} \wedge dx^{\nu} \quad ,  \quad  
S_{\accentset{\mbox{\huge .}}{A}\accentset{\mbox{\huge .}}{B}}= \frac{1}{2} S_{\enspace \enspace \mu \nu}^{ \accentset{\mbox{\huge .}}{A}\accentset{\mbox{\huge .}}{B}}dx^{\mu} \wedge dx^{\nu}
\ee
with the obvious rules for raising and lowering indices
\be
\label{215bd}
S^{\mu \nu}_{\enspace  AB}= \eta^{\mu \alpha} \eta^{\nu \beta }\epsilon_{AC} \epsilon_{BD} S_{\alpha \beta}^{\enspace CD} \quad, \quad
S^{\mu \nu}_{\enspace \accentset{\mbox{\huge .}}{A}\accentset{\mbox{\huge .}}{B}}= \eta^{\mu \alpha} \eta^{\nu \beta }
\epsilon_{\accentset{\mbox{\huge .}}{A}\accentset{\mbox{\huge .}}{C}} \epsilon_{\accentset{\mbox{\huge .}}{B}\accentset{\mbox{\huge .}}{D}}  S_{\alpha \beta}^{\enspace \accentset{\mbox{\huge .}}{C}\accentset{\mbox{\huge .}}{D}}.
\ee
One can check easily the following identities
\begin{subequations}\label{215cd}
\begin{align}
   S_{\mu \nu}^{\enspace AB}S^{\mu \nu}_{\enspace \accentset{\mbox{\huge .}}{C}\accentset{\mbox{\huge .}}{D}}&=0=
    S_{\mu \nu}^{\enspace \accentset{\mbox{\huge .}}{A}\accentset{\mbox{\huge .}}{B}}S^{\mu \nu}_{\enspace CD} ,
   \label{215acd} \\
   S_{\mu \nu}^{\enspace AB}S^{\mu \nu}_{\enspace CD}&=4\left( \delta^A_C\, \delta^B_D + \delta^A_D\, \delta^B_C \right),  \label{215bcd} \\
   S_{\mu \nu}^{\enspace \accentset{\mbox{\huge .}}{A}\accentset{\mbox{\huge .}}{B}}S^{\mu \nu}_{\enspace \accentset{\mbox{\huge .}}{C}\accentset{\mbox{\huge .}}{D}}&=4 \left(\delta^{\accentset{\mbox{\huge .}}{A}}_{\accentset{\mbox{\huge .}}{C}} \, \delta^{\accentset{\mbox{\huge .}}{B}}_{\accentset{\mbox{\huge .}}{D}}+
   \delta^{\accentset{\mbox{\huge .}}{A}}_{\accentset{\mbox{\huge .}}{D}} \, \delta^{\accentset{\mbox{\huge .}}{B}}_{\accentset{\mbox{\huge .}}{C}}
   \right).  \label{125ccd}
\end{align}
\end{subequations}
Applying (\ref{215cd}) to (\ref{215}) we obtain
\be
\label{216}
f_{AB}= \frac{1}{4} \mathcal{F}_{\mu \nu} S^{\mu \nu}_{\enspace AB} \quad , \quad 
f_{\accentset{\mbox{\huge .}}{A}\accentset{\mbox{\huge .}}{B}}= \frac{1}{4} \mathcal{F}_{\mu \nu} S^{\mu \nu}_{\enspace \accentset{\mbox{\huge .}}{A}\accentset{\mbox{\huge .}}{B}}.
\ee
Straightforward calculations lead to relations
\begin{align}\label{218}
  f_{\accentset{\mbox{\huge .}}{1}\accentset{\mbox{\huge .}}{1}} &= \frac{1}{\sqrt{2}}\left(F_{1}-iF_{2}\right)=(f_{11})^{ \ast}, \nonumber \\
  f_{\accentset{\mbox{\huge .}}{1}\accentset{\mbox{\huge .}}{2}} &= f_{\accentset{\mbox{\huge .}}{2}\accentset{\mbox{\huge .}}{1}}=-\frac{1}{\sqrt{2}}F_{3}=(f_{12})^{ \ast}=(f_{21})^{ \ast}, \nonumber \\
  f_{\accentset{\mbox{\huge .}}{2}\accentset{\mbox{\huge .}}{2}} &= -\frac{1}{\sqrt{2}}\left(F_{1}+iF_{2}\right)=( f_{22})^{ \ast}
\end{align}
where $(F_{1}, F_{2},F_{3})=\vec{F}$ is the \textit{Riemann -- Silberstein vector} \cite{Bialynicki1994, Bialynicki1996, Silberstein1907, Weber1901}
\begin{equation}\label{219}
  \vec{F}:= \frac{1}{\sqrt{2}}\left(\vec{\mathcal{E}}+i\vec{\mathcal{B}}\right).
\end{equation}

Equations (\ref{218}) can be rewritten in a compact form as
\be
\label{220}
 f_{\accentset{\mbox{\huge .}}{A}\accentset{\mbox{\huge .}}{B}} = i\, \Phi^j_{\; \accentset{\mbox{\huge .}}{A}\accentset{\mbox{\huge .}}{B}}\,F_j= (f_{AB})^{\ast} \quad, \quad j=1,2,3
\ee
with $\Phi^j_{\; \accentset{\mbox{\huge .}}{A}\accentset{\mbox{\huge .}}{B}}$ defined by $2 \times 2$ matrices
\be
\label{221}
(\Phi^1_{\; \accentset{\mbox{\huge .}}{A}\accentset{\mbox{\huge .}}{B}}):= \frac{1}{\sqrt{2}}
\left( \begin{array}{cc}
                                      -i & 0 \\
                                      0 & i \\
                                    \end{array}
                                  \right)\, ,\,
 (\Phi^2_{\; \accentset{\mbox{\huge .}}{A}\accentset{\mbox{\huge .}}{B}}):=  \frac{1}{\sqrt{2}}                               
                                  \left( \begin{array}{cc}
                                      -1 & 0 \\
                                      0 &-1 \\
                                    \end{array}
                                  \right)\, , \,
  (\Phi^3_{\; \accentset{\mbox{\huge .}}{A}\accentset{\mbox{\huge .}}{B}}):=  \frac{1}{\sqrt{2}}                              
                                  \left( \begin{array}{cc}
                                      0 & i \\
                                      i &0 \\
                                    \end{array}
                                  \right).
\ee
Rising spinorial indices according to the rule (\ref{27}) we obtain that
\be
\label{223}
(\Phi^{1 \accentset{\mbox{\huge .}}{A}\accentset{\mbox{\huge .}}{B}}):= \frac{1}{\sqrt{2}}
\left( \begin{array}{cc}
                                      i & 0 \\
                                      0 & -i \\
                                    \end{array}
                                  \right)\, ,\,
 (\Phi^{2 \accentset{\mbox{\huge .}}{A}\accentset{\mbox{\huge .}}{B}}):=  \frac{1}{\sqrt{2}}                               
                                  \left( \begin{array}{cc}
                                      -1 & 0 \\
                                      0 &-1 \\
                                    \end{array}
                                  \right)\, , \,
  (\Phi^{3 \accentset{\mbox{\huge .}}{A}\accentset{\mbox{\huge .}}{B}}):=  \frac{1}{\sqrt{2}}                              
                                  \left( \begin{array}{cc}
                                      0 & i \\
                                      i &0 \\
                                    \end{array}
                                  \right).
\ee
Thus one immediately finds that
\be
\label{222}
\Phi^j_{\; \accentset{\mbox{\huge .}}{A}\accentset{\mbox{\huge .}}{B}}
\Phi^{k \accentset{\mbox{\huge .}}{A}\accentset{\mbox{\huge .}}{B}}= \delta^{jk} \quad, \quad
j,k=1,2,3.
\ee
From (\ref{220}) and (\ref{222}) we get
\be
\label{224}
F_j= -i \, \Phi_j^{\; \accentset{\mbox{\huge .}}{A}\accentset{\mbox{\huge .}}{B}}f_{ \accentset{\mbox{\huge .}}{A}\accentset{\mbox{\huge .}}{B}}
\ee
under the convention that the Latin indices $j,k=1,2,3$ are to be manipulated with the use of the Kronecker delta so $\Phi_j^{\; \accentset{\mbox{\huge .}}{A}\accentset{\mbox{\huge .}}{B}}=\Phi^{j \accentset{\mbox{\huge .}}{A}\accentset{\mbox{\huge .}}{B}}.$

Inserting Eq. (\ref{224}) back into (\ref{220}) one quickly finds the relation
\be
\label{225}
\Phi_j^{\; \accentset{\mbox{\huge .}}{A}\accentset{\mbox{\huge .}}{B}}
\Phi^{j}_{\; \accentset{\mbox{\huge .}}{C}\accentset{\mbox{\huge .}}{D}}= \frac{1}{2}
\left(\delta^{\accentset{\mbox{\huge .}}{A}}_{\accentset{\mbox{\huge .}}{C}} \, \delta^{\accentset{\mbox{\huge .}}{B}}_{\accentset{\mbox{\huge .}}{D}}+
   \delta^{\accentset{\mbox{\huge .}}{A}}_{\accentset{\mbox{\huge .}}{D}} \, \delta^{\accentset{\mbox{\huge .}}{B}}_{\accentset{\mbox{\huge .}}{C}}
\right).
\ee
Observe that the scalar product
\be
\label{226}
\vec{F} \cdot \vec{F} = \frac{1}{2} \left( {\vec{\mathcal E}}^{\,2} - {\vec{\mathcal B}}^{\,2} + 2i \,\vec{\mathcal E} \cdot \vec{\mathcal B}
\right)
\ee 
is invariant under the {\it proper ortochronous Lorentz group} $L^{\shneg}_{+}.$ Consequently, if $(\Lambda^{\mu}_{\enspace \nu}) \in L^{\shneg}_{+}$ defines a transformation from an inertial system of frames $K$ to another inertial system $K'$ and $(l^{\accentset{\mbox{\huge .}}{A}}_{\enspace \accentset{\mbox{\huge .}}{B}}), - (l^{\accentset{\mbox{\huge .}}{A}}_{\enspace \accentset{\mbox{\huge .}}{B}}) \in SL(2, {\mathbb C}) $ are two possible representatives of transformation $(\Lambda^{\mu}_{\enspace \nu})$ in the spinor  representation,  then we have
\be
\label{227}
F'_j= -i \, \Phi_{j \accentset{\mbox{\huge .}}{A}\accentset{\mbox{\huge .}}{B}}f'^{ \,\accentset{\mbox{\huge .}}{A}\accentset{\mbox{\huge .}}{B}}=  -i \, \Phi_{j \accentset{\mbox{\huge .}}{A}\accentset{\mbox{\huge .}}{B}}
l^{\accentset{\mbox{\huge .}}{A}}_{\enspace \accentset{\mbox{\huge .}}{C}}
l^{\accentset{\mbox{\huge .}}{B}}_{\enspace \accentset{\mbox{\huge .}}{D}}
f^{ \,\accentset{\mbox{\huge .}}{C}\accentset{\mbox{\huge .}}{D}}
\ee
and
\be
\label{228}
{F'_{1}}^{2}+{F'_{2}}^{2}+{F'_{3}}^{2}=
{F_{1}}^{2}+{F_{2}}^{2}+{F_{3}}^{2}=\vec{F} \cdot \vec{F}
\ee
where the prime `` ${}'\;$'' corresponds to objects in the inertial system $K'.$ From (\ref{228}) it follows that there exists an orthogonal complex $3 \times 3$ matrix $(t^{\,j}_{\enspace k}) \in O(3 ; {\mathbb C})$ such that
\be
\label{229}
F'^j =t^j_{\; k} F^k \quad; \quad F'^j \equiv F'_j \quad, \quad F^k \equiv F_k.
\ee
Substituting (\ref{220}) into the right -- hand side and (\ref{229}) into the left -- hand side of (\ref{227}) one obtains
\be
\label{230}
t^j_{\; k}=  \Phi^j_{ \accentset{\mbox{\huge .}}{A}\accentset{\mbox{\huge .}}{B}}\,
l^{\accentset{\mbox{\huge .}}{A}}_{\enspace \accentset{\mbox{\huge .}}{C}}\,
l^{\accentset{\mbox{\huge .}}{B}}_{\enspace \accentset{\mbox{\huge .}}{D}}\,
\Phi_k^{ \accentset{\mbox{\huge .}}{C}\accentset{\mbox{\huge .}}{D}}.
\ee
Employing then (\ref{225}) we get the inverse relation
\be
\label{231}
\frac{1}{2} \left(
l^{\accentset{\mbox{\huge .}}{A}}_{\enspace \accentset{\mbox{\huge .}}{C}}
l^{\accentset{\mbox{\huge .}}{B}}_{\enspace \accentset{\mbox{\huge .}}{D}} +
l^{\accentset{\mbox{\huge .}}{A}}_{\enspace \accentset{\mbox{\huge .}}{D}}
l^{\accentset{\mbox{\huge .}}{B}}_{\enspace \accentset{\mbox{\huge .}}{C}}
 \right)=
 \Phi_j^{ \accentset{\mbox{\huge .}}{A}\accentset{\mbox{\huge .}}{B}}
 t^j_{\; k}
 \Phi^k_{ \accentset{\mbox{\huge .}}{C}\accentset{\mbox{\huge .}}{D}}.
\ee
From (\ref{230}), using also (\ref{221}) and (\ref{223}), one shows that 
\be
\label{232}
\det (t^j_{\; k})=1 \quad \Longrightarrow \quad (t^j_{\; k}) \in SO(3; {\mathbb C}).
\ee
Moreover, a simple analysis of Eqs. (\ref{230}) and (\ref{231}) leads to conclusion that matrices $ \left(l^{\accentset{\mbox{\huge .}}{A}}_{\enspace \accentset{\mbox{\huge .}}{B}} \right)$ and 
$- \left( l^{\accentset{\mbox{\huge .}}{A}}_{\enspace \accentset{\mbox{\huge .}}{B}} \right)$ define the same matrix $(t^j_{\; k})$ and conversely, every matrix $(t^j_{\; k})$ defines the $SL(2;{\mathbb C})$ matrix $ \left(l^{\accentset{\mbox{\huge .}}{A}}_{\enspace \accentset{\mbox{\huge .}}{B}} \right)$ up to the sign. Therefore, formulae (\ref{230}) and (\ref{231}) realise a group isomorphism
\be
\label{233}
SL(2;\mathbb{C}) \mbox{\Large $/$} \mathbb{Z}_2 \cong SO(3;\mathbb{C})
\ee
where $\mathbb{Z}_2= \{+1,-1 \}$ is the cyclic group.

Since $SL(2;\mathbb{C}) \mbox{\Large $/$} \mathbb{Z}_2 \cong L^{\shneg}_{+}$ one obtains an isomorphism
\be
\label{234}
L^{\shneg}_{+} \cong  SO(3;\mathbb{C}).
\ee
Thus we arrive at a principal bundle isomorphism
\be
\label{235}
\left( M_4 \times L^{\shneg}_{+}, M_4, \Pi_1 \right) \cong
\left( M_4 \times SO(3; \mathbb{C}), M_4, \Pi_2 \right) 
\ee
where $M_4$ is the Minkowski spacetime and 
\[
\Pi_1:M_4 \times L^{\shneg}_{+} \rightarrow  M_4 \quad,\quad
 \Pi_2:M_4 \times SO(3; \mathbb{C}) \rightarrow M_4
 \]
  are natural projections on $M_4.$ Principal fibre bundle $\left( M_4 \times SO(3; \mathbb{C}), M_4, \Pi_2 \right) $ can be considered as a reduced bundle of all linear bases in vector space ${\mathbb C}^3$ over $M_4$ given by a reduction of Lie group $GL(3;\mathbb C)$ to $SO(3; \mathbb{C}).$ With such an identification one can construct vector bundles associated with  principle bundle 
$\left( M_4 \times SO(3; \mathbb{C}), M_4, \Pi_2 \right). $ 
Sections of those vector bundles are vector and tensor fields on $M_4$ which transform according to a suitable representation of  $SO(3; \mathbb{C}).$

From (\ref{220}) and (\ref{224}) one concludes that any vector bundle of $SO(3; \mathbb{C})$
vectors is isomorphic to the vector bundle of dotted symmetric spinors of the second rank. Of course, {\it mutatis mutandi}, similar construction can be done for the undotted symmetric spinors of the second rank. In this construction Eqs. (\ref{220}) and (\ref{224}) take the form
\be
\label{235}
f_{AB}=-i \,\Phi^{\ast j}_{\enspace AB} F^{\ast}_j
\ee
and
\be
\label{236}
F^{\ast}_j= i \, \Phi^{\ast  AB}_{j} f_{AB}
\ee
where
\be
\label{237}
\Phi^{\ast j}_{\enspace AB}:= \left( \Phi^j_{ \;\accentset{\mbox{\huge .}}{A}\accentset{\mbox{\huge .}}{B}}\right)^{\ast}.
\ee
According to the Pleba\'{n}ski's terminology \cite{Plebanski1980} the formalism described above as founded on the $SO(3; \mathbb{C})$ group we call the {\it helicity formalism} \footnote{In his notes \cite{Plebanski1980} Jerzy Pleba\'{n}ski developed the helicity formalism in  detail  for all real and complex Riemannian $4$ -- dimensional structures.}.

From relations (\ref{220}) or (\ref{224}) one quickly infers that $\Phi^j_{ \;\accentset{\mbox{\huge .}}{A}\accentset{\mbox{\huge .}}{B}}$ is a {\it spin -- $SO(3; \mathbb{C})$ vector.}

In spinorial formalism the Maxwell equations in empty space read
\be
\label{238}
\partial^{A \accentset{\mbox{\huge .}}{B} } f_ { \accentset{\mbox{\huge .}}{B}\accentset{\mbox{\huge .}}{C}}=0
\Longleftrightarrow 
\partial^{B \accentset{\mbox{\huge .}}{A} } f_ { BC}=0
\ee
where
\be
\label{239}
\partial^{A \accentset{\mbox{\huge .}}{B} }:= g^{\mu A \accentset{\mbox{\huge .}}{B} }\partial_{\mu}=
\left(\partial^{B \accentset{\mbox{\huge .}}{A} } \right)^{\ast}.
\ee
Under convention used in the present paper it is convenient to deal with the form of Maxwell equations as written on the left -- hand side of (\ref{238}).

Inserting relations (\ref{239}) with (\ref{24}) into the Maxwell equations and employing symmetry of $ f_ { \accentset{\mbox{\huge .}}{A}\accentset{\mbox{\huge .}}{B}}$ one concludes that the Maxwell equations split into the evolution matrix equation
\be
\label{240}
\partial_t {\mathbf f}= -c \left(\vec{\mathcal S}' \cdot \vec{\nabla} \right) {\mathbf f}
\ee
where
\setlength{\extrarowheight}{3pt}
\[
{\mathbf f}:=
 \left(
    \begin{array}{c} 
      f_{\accentset{\mbox{\huge .}}{1}\accentset{\mbox{\huge .}}{1}} \\ 
      \sqrt{2}\,f_{\accentset{\mbox{\huge .}}{1}\accentset{\mbox{\huge .}}{2}} \\
      f_{\accentset{\mbox{\huge .}}{2}\accentset{\mbox{\huge .}}{2}} 
    \end{array}
  \right) \quad , \quad \vec{\nabla}= (\partial_1, \partial_2, \partial_3)
\]
and $\vec{\mathcal S}'= ({\mathcal S}'_1, {\mathcal S}'_2, {\mathcal S}'_3) $ with
\setlength{\extrarowheight}{0pt}
\be
\label{241}
{\mathcal S}'_1=  \frac{1}{\sqrt{2}}
\left(
    \begin{array}{ccc}
      0 & 1 &0 \\
      1 & 0 &1 \\
     0 & 1 & 0
    \end{array}
  \right)
  \; , \;
{\mathcal S}'_2= \frac{1}{\sqrt{2}}
\left(
    \begin{array}{ccc}
      0 & -i &0 \\
      i & 0 &-i \\
     0 & i & 0
    \end{array}
  \right)
   \; , \;
{\mathcal S}'_3= \frac{1}{\sqrt{2}}
\left(
    \begin{array}{ccc}
      1 & 0 &0 \\
      0 & 0 &0 \\
     0 & 0 & -1
    \end{array}
  \right)
\ee
and the constraint equation
\be
\label{242}
\partial_j \left( \Phi^{j \,\accentset{\mbox{\huge .}}{A}\accentset{\mbox{\huge .}}{B}} f_{\accentset{\mbox{\huge .}}{A}\accentset{\mbox{\huge .}}{B}}\right)=0.
\ee
Writing the Riemann -- Silberstein vector $\vec{F}$ in a matrix form
\be
\label{243}
{\mathbf F}:= \left(
    \begin{array}{c}
      F_1 \\
      F_2 \\
     F_3
    \end{array}
  \right)
\ee
we can rewrite relations (\ref{220}) and (\ref{224}) in the form
\be
\label{244}
{\mathbf f}= {\mathcal U} \cdot {\mathbf F} \quad {\mbox and} \quad 
{\mathbf F}= {\mathcal U}^{\dagger} \cdot {\mathbf f}
\ee
where
\setlength{\extrarowheight}{3pt}
\be
\label{245}
{\mathcal U}= \frac{1}{\sqrt{2}}
\left(
    \begin{array}{ccc}
      1 & -i &0 \\
      0 & 0 & - \sqrt{2}  \\
     -1 & -i & 0
    \end{array}
  \right) = i
\left(
    \begin{array}{ccc}
      \Phi^{1}_{ \,\accentset{\mbox{\huge .}}{1}\accentset{\mbox{\huge .}}{1}} & \Phi^{2}_{ \,\accentset{\mbox{\huge .}}{1}\accentset{\mbox{\huge .}}{1}} &\Phi^{3}_{ \,\accentset{\mbox{\huge .}}{1}\accentset{\mbox{\huge .}}{1}} \\
    \sqrt{2}  \Phi^{1}_{ \,\accentset{\mbox{\huge .}}{1}\accentset{\mbox{\huge .}}{2}} &  \sqrt{2} \Phi^{2}_{ \,\accentset{\mbox{\huge .}}{1}\accentset{\mbox{\huge .}}{2}} &\sqrt{2} \Phi^{3}_{ \,\accentset{\mbox{\huge .}}{1}\accentset{\mbox{\huge .}}{2}} \\
   \Phi^{1}_{ \,\accentset{\mbox{\huge .}}{2}\accentset{\mbox{\huge .}}{2}} & \Phi^{2}_{ \,\accentset{\mbox{\huge .}}{2}\accentset{\mbox{\huge .}}{2}} &\Phi^{3}_{ \,\accentset{\mbox{\huge .}}{2}\accentset{\mbox{\huge .}}{2}}   
    \end{array}
  \right)
\ee
\setlength{\extrarowheight}{0pt}
is a unitary $3 \times 3$ matrix. Then ${\mathcal U}^{\dagger}$ is a Hermitian conjugation of 
${\mathcal U}$
\be
\label{246}
{\mathcal U}^{\dagger}= \frac{1}{\sqrt{2}}
\left(
    \begin{array}{ccc}
      1 & 0 &-1 \\
      i & 0 & i  \\
     0 & -\sqrt{2} & 0
    \end{array}
  \right).
\ee
Multiplying both sides of Eq. (\ref{240}) by ${\mathcal U}^{\dagger}$ on the left -- hand side and using (\ref{244}) one gets the Maxwell evolution equation in terms of the Riemann -- Silberstein vector as
\be
\label{247}
\partial_t {\mathbf F}= -c \left(\vec{\mathcal S} \cdot \vec{\nabla} \right) {\mathbf F},
\ee
where
\[
\vec{\mathcal S}= ({\mathcal S}_1, {\mathcal S}_2, {\mathcal S}_3):= {\mathcal U}^{\dagger}\vec{\mathcal S}' {\mathcal U}
\]
and
\be
\label{248}
{\mathcal S}_1=  
\left(
    \begin{array}{ccc}
      0 & 0 &0 \\
      0 & 0 &-i \\
     0 & i & 0
    \end{array}
  \right)
  \; , \;
{\mathcal S}_2= 
\left(
    \begin{array}{ccc}
      0 & 0 &i \\
      0 & 0 &0 \\
     -i & 0 & 0
    \end{array}
  \right)
   \; , \;
{\mathcal S}_3= 
\left(
    \begin{array}{ccc}
      0 & -i &0 \\
      i & 0 &0 \\
     0 & 0 & 0
    \end{array}
  \right).
\ee
In a compact form matrices ${\mathcal S}_j$ can be written with the use of the $3$ -- D Levi -- Civita symbol 
\be
\label{249}
({\mathcal S}_{j})_{ kl}= -i \epsilon_{jkl} \quad , \quad j,k,l=1,2,3.
\ee

Then comparing (\ref{242}) with (\ref{224}) one quickly finds that the constraint equation (\ref{242}) in terms of ${\mathbf F}$ reads
\be
\label{250}
\sum_{j=1}^3
\partial_j {\mathbf F}_j=0.
\ee

Matrices ${\mathcal S}_{j}, \; j=1,2,3$ are the well known {\it spin -- $1$ matrices}. They satisfy the following commutation relations
\be
\label{251}
{\mathcal S}_{j} \cdot {\mathcal S}_{k} - {\mathcal S}_{k} \cdot {\mathcal S}_{j}= i \epsilon_{jkl} {\mathcal S}_{l}.
\ee
From (\ref{248}) we infer that the system of matrices $\{{\mathcal S}_{1},{\mathcal S}_{2},{\mathcal S}_{3} \}$ is unitary equivalent to the system  $\{{\mathcal S}'_{1},{\mathcal S}'_{2},{\mathcal S}'_{3} \}.$ Hence matrices ${\mathcal S}'_{j}$ are also the spin -- $1$ matrices in another representation. 

They fulfill the same commutation relation (\ref{251}) as  matrices ${\mathcal S}_{j}$ do i.e. 
\be
\label{252}
{\mathcal S}'_{j} \cdot {\mathcal S}'_{k} - {\mathcal S}'_{k} \cdot {\mathcal S}'_{j}= i \epsilon_{jkl} {\mathcal S}'_{l}.
\ee
One can show that the constraint equation (\ref{250}) is equivalent to expression \cite{Bialynicki1996, Pryce1948}
\be
\label{253}
\left( \vec{\mathcal S}  \cdot \vec{\nabla}\right) \cdot {\mathcal S}_{j} {\mathbf F} = \partial_j {\mathbf F} 
\ee
for any $j$. Acting on both sides of (\ref{253}) with $ \partial_j$ and summing over $j$ we get 
\cite{Bialynicki1996, Pryce1948}
\be
\label{254}
\left( \vec{\mathcal S}  \cdot \vec{\nabla}\right)^2 \cdot  {\mathbf F} = \Delta {\mathbf F}. 
\ee
Of course, Eqs. (\ref{253}) and (\ref{254}) can be equivalently rewritten in terms of matrix ${\mathbf f}$ as
\be
\label{255}
\left( \vec{\mathcal S}'  \cdot \vec{\nabla}\right) \cdot {\mathcal S}'_{j} \,{\mathbf f} = \partial_j {\mathbf f}
\ee
and
\be
\label{256}
\left( \vec{\mathcal S}'  \cdot \vec{\nabla}\right)^2 \cdot  {\mathbf f} = \Delta {\mathbf f}. 
\ee
Any solution of Eq. (\ref{247}) can be expanded in the plane wave solutions
\be
\label{257}
{\mathbf F}(\vec{x},t)= \int_{{\mathbb R}^3}\frac{d^3 k}{(2 \pi)^3} 
\left[ {\bf a}_{+}(\vec{k}) \exp \{i(\vec{k} \cdot \vec{x}-\omega_k t) \}
+ {\bf a}_{-}(\vec{k}) \exp \{-i(\vec{k} \cdot \vec{x}-\omega_k t) \}
\right], 
\ee
where 
\be
\label{258}
\omega_k= c |\vec{k}| \quad , \quad {\bf a}_{\pm} (\vec{k})= 
\left( \begin{array}{c}
a_{1 \pm} (\vec{k}) \\
a_{2 \pm} (\vec{k}) \\
a_{3 \pm} (\vec{k}) 
\end{array}
\right).
\ee
Inserting (\ref{257}) into (\ref{247}) one arrives at conclusions
\be
\label{259}
\left( \vec{\mathcal S} \cdot \vec{k} \right) {\bf a}_{\pm} (\vec{k}) = |\vec{k}| \, {\bf a}_{\pm} (\vec{k})
\ee
or, by (\ref{249}), to the system of three conditions
\be
\label{260}
\sum_{l,m=1}^3
i \,\epsilon_{jlm}\, k_l \,a_{m \pm}(\vec{k}) =  |\vec{k}| a_{j \pm} (\vec{k}).
\ee
Then  constraint (\ref{250}) under (\ref{249}) gives
\be
\label{261}
\sum_{j=1}^3
k_j  a_{j \pm} (\vec{k})=0.
\ee
From (\ref{252}) (or from (\ref{251}) with (\ref{248})) we find 
\be
\label{262}
{\bf a}_{+}^{\rm T}(\vec{k}) \cdot {\bf a}_{+}(\vec{k})=0=
{\bf a}_{-}^{\rm T}(\vec{k}) \cdot {\bf a}_{-}(\vec{k}).
\ee
Elementary analysis of Eqs. (\ref{260}) and (\ref{261}) leads to the conclusion that 
${\bf a}_{\pm}(\vec{k})$
can be written in the form
\be
\label{263}
{\bf a}_{\pm}(\vec{k})= \alpha_{\pm}(\vec{k}) {\bf e} (\vec{k}),
\ee
where $\alpha_{\pm}(\vec{k})$ are some functions of $\vec{k}$ and
\[
{\bf e} (\vec{k})= \left( \begin{array}{c}
e_1(\vec{k}) \\
e_2(\vec{k}) \\
e_3(\vec{k})
\end{array}
\right)
\]
is a solution  of the system of equations
\begin{subequations}\label{264}
\begin{align}
   \left( \vec{\mathcal S} \cdot \vec{k} \right) {\bf e}(\vec{k}) = |\vec{k}| \, {\bf e} (\vec{k}) , \label{264a}\\
  \sum_{j=1}^3
k_j  e_{j } (\vec{k})=0\label{264b}
\end{align}
\end{subequations}
fulfilling the normalisation condition
\be
\label{265}
{\bf e}^{\dagger}(\vec{k}) \cdot {\bf e}(\vec{k})=1.
\ee
By \eqref{262} one has 
\be
\label{266}
{\bf e}^{\rm T}(\vec{k}) \cdot {\bf e}(\vec{k})=0.
\ee
Therefore ${\bf e}(\vec{k})$ can be written as
\be
\label{267}
{\bf e}(\vec{k})= \frac{1}{\sqrt{2}} \left({\bf m}(\vec{k}) + i {\bf n}(\vec{k}) \right),
\ee
where
\[
{\bf m}(\vec{k})= \left(\begin{array}{c}
m_1(\vec{k}) \\
m_2(\vec{k}) \\
m_3(\vec{k}) 
\end{array}
\right) 
\quad , \quad
{\bf n}(\vec{k})= \left(\begin{array}{c}
n_1(\vec{k}) \\
n_2(\vec{k}) \\
n_3(\vec{k}) 
\end{array}
\right) 
\]
are real $1$  -- column matrices satisfying the following conditions
\[
{\bf m}^{\rm T}(\vec{k}) \cdot {\bf m}(\vec{k})=1 \; , \;
{\bf n}^{\rm T}(\vec{k}) \cdot {\bf n}(\vec{k})=1 \; ,\;
{\bf m}^{\rm T}(\vec{k}) \cdot {\bf n}(\vec{k})=0,
\]
\be
\label{268}
\sum_{j=1}^3 k_j m_j(\vec{k})=0 \; , \; \sum_{j=1}^3 k_j n_j(\vec{k})=0 \; , \; 
\sum_{l,r=1}^3\epsilon_{jlr} m_l (\vec{k}) n_r (\vec{k}) = \frac{k_j}{|\vec{k}|}.
\ee
From \eqref{264} and \eqref{248} one obtains
\begin{subequations}
\label{269}
\begin{align}
 \left( \vec{\mathcal S} \cdot \vec{k} \right) {\bf e}^{\ast}(\vec{k}) =- |\vec{k}| \, {\bf e}^{\ast} (\vec{k}),  \label{269a}\\
  \sum_{j=1}^3
k_j  e_{j }^{\ast} (\vec{k})=0.\label{269b}
\end{align}
\end{subequations}
Substituting \eqref{263} into \eqref{257} we can see that
\be
\label{270}
{\mathbf F}(\vec{x},t)= 
\int_{{\mathbb R}^3}\frac{d^3 k}{(2 \pi)^3} \,{\bf e}(\vec{k})
\left[ {\alpha}_{+}(\vec{k}) \exp \{i(\vec{k} \cdot \vec{x}-\omega_k t) \}
+ {\alpha}_{-}(\vec{k}) \exp \{-i(\vec{k} \cdot \vec{x}-\omega_k t) \}
\right]. 
\ee
Multiplying both sides of \eqref{270} on the left -- hand  by ${\mathcal U}$ and employing (\ref{244}) one gets
\be
\label{271}
{\mathbf f}(\vec{x},t)= 
\int_{{\mathbb R}^3}\frac{d^3 k}{(2 \pi)^3}\, {\mathcal U} \cdot {\bf e}(\vec{k})
\left[ {\alpha}_{+}(\vec{k}) \exp \{i(\vec{k} \cdot \vec{x}-\omega_k t) \}
+ {\alpha}_{-}(\vec{k}) \exp \{-i(\vec{k} \cdot \vec{x}-\omega_k t) \}
\right]. 
\ee
Matrix ${\bf e}(\vec{k})$ is determined by ${\mathbf F}(\vec{x},t)$ up to a phase factor
\be
\label{272}
{\bf e}(\vec{k}) \longrightarrow \exp \{i \beta (\vec{k}) \} {\bf e}(\vec{k})
\quad , \quad
 \beta^{\ast}(\vec{k})=\beta (\vec{k}).
\ee
Careful analysis of \eqref{264} and \eqref{269} shows that one can choose $\beta (\vec{k})$
so that
\be
\label{273}
 {\bf e}(-\vec{k})=  {\bf e}^{\ast}(\vec{k}).
\ee
From now we assume that condition \eqref{273} holds true.

Finally, let us write down formulae for energy $E,$ momentum $\vec{P},$ angular momentum $\vec{M}$ and moment of energy $\vec{N}$ of the Maxwell electromagnetic field \cite{Bialynicki1996, Bialynicki1975}
\begin{subequations}
\label{274}
\begin{align}
E&= \int_{{\mathbb R}^3} d^3x \,\vec{ F}^{\ast} \cdot  \vec{ F}=
\int_{{\mathbb R}^3}\frac{d^3 k}{(2 \pi)^3} \left( | {\alpha}_{+}(\vec{k})|^2
+ {\alpha}_{-}(\vec{k})|^2
\right)= 
\int_{{\mathbb R}^3} d^3x \,{\mathbf f}^{\dagger} \cdot  {\mathbf f}, \label{274a}\\
\vec{P}&= \frac{1}{ic }
\int_{{\mathbb R}^3} d^3x \,\vec{ F}^{\ast} \times  \vec{ F}=
\int_{{\mathbb R}^3}\frac{d^3 k}{(2 \pi)^3}
\frac{\vec{k}}{\omega_k}
 \left( | {\alpha}_{+}(\vec{k})|^2
+ {\alpha}_{-}(\vec{k})|^2
\right), \label{274b}\\
\vec{M}&=  \frac{1}{ic }
\int_{{\mathbb R}^3} d^3x \, \vec{x} \times \left(\vec{ F}^{\ast} \times  \vec{ F}\right), \label{274c}\\
\vec{N}&= \int_{{\mathbb R}^3} d^3x \,\vec{x} \left(\vec{ F}^{\ast} \cdot  \vec{ F} \right). \label{274d}
\end{align}
\end{subequations}

\section{The wave function of photon}
\label{sec3}
\setcounter{equation}{0}

It is a trivial statement that photon is, {\it par exellence}, a relativistic particle. Consequently, a natural theory describing any system of photons is relativistic quantum field theory. Nevertheless several authors \cite{Bialynicki1994, Bialynicki1996, Bialynicki2006, Sipe1995, Bialynicki2017, Chandrasekar2012, Landau1930, Oppenheimer1931, Moliere1950, Good1957, Cook1982, Inagaki1994, Hawton1999a}, in analogy to the case of the Dirac equation which provides quantum mechanical but not quantum field description of electron, are trying to introduce a wave function of photon and to construct quantum mechanics of this particle. 

In the present section we consider a version of quantum   mechanical approach to a single photon system developed by I.  Bia\l ynicki -- Birula \cite{Bialynicki1994, Bialynicki1996} and J. E. Sipe \cite{Sipe1995} in their distinguished works.

First, let us multiply both sides of \eqref{247} by $i \hbar.$ This leads to 
\be
\label{31}
i \hbar \,\partial_t {\mathbf F}= c \left(\vec{\mathcal S} \cdot \widehat{\vec{p}} \,\right) {\mathbf F},
\ee
where $\widehat{\vec{p}}= -i\hbar \vec{\nabla}= \left(-i\hbar \partial_1, -i\hbar \partial_2,
-i\hbar \partial_3 \right)$ is the momentum operator. 

In the same way from \eqref{240} one obtains
\be
\label{32}
i\hbar \,\partial_t {\mathbf f}= c \left(\vec{\mathcal S}' \cdot  \widehat{\vec{p}} \right) {\mathbf f}.
\ee
At first glance Eq. \eqref{31} is similar to the Dirac equation with an obvious additional assumption that now we are dealing not with an electron but with a photon which is \\ a   spin -- $1$ massless particle. This observation suggests that Eq. \eqref{31} is the right quantum mechanical relativistic equation for a photon and ${\mathbf F}= {\mathbf F}(\vec{x},t)$ subject to additional condition \eqref{250} is the photon wave function. 

However, this conclusion is incorrect since Eq. \eqref{31} considered as a quantum evolution equation admits solutions with negative energy
\[
 {\bf e}(\vec{k})
 {\alpha}_{-}(\vec{k}) \exp \{-i(\vec{k} \cdot \vec{x}-\omega_k t) \} \quad , \quad
 \vec{k} = \frac{\vec{p}}{\hbar}
\]
(see \eqref{264} and \eqref{270}).
In the case of the Dirac equation the solutions with negative energies are interpreted as the ones representing the antiparticle. However the antiparticle of photon is the photon itself. Therefore in the photon quantum mechanics the solutions with negative energies are unphysical. It means that some modification of formula \eqref{31} is required.

First note that the {\it helicity operator of photon} in the momentum representation reads
\be
\label{33}
\widehat{\mathbf \Sigma} = \frac{\vec{\mathcal S} \cdot \vec{p}}{|\vec{p}|}= 
 \frac{\vec{\mathcal S} \cdot \vec{k}}{|\vec{k}|} \quad , \quad \vec{p}= \hbar \vec{k}
\ee
and its eigenvalues are $\lambda= \pm 1.$ A general wave function is a superposition of $(+1)$ -- helicity states and $(-1)$ -- helicity states. Quick look at Eqs. \eqref{264}, \eqref{269} and \eqref{270} with taking into account that only states of positive energy have physical meaning, lead to conclusion that the \textit{photon wave function} has a form
\be
\label{34}
{\bf \Psi}(\vec{x},t)= \sqrt{\hbar c} 
\int_{{\mathbb R}^3}\frac{d^3 k}{(2 \pi)^3} \,  \left[ {\bf e}(\vec{k})
\alpha (\vec{k}, +1) +{\bf e}^{\ast}(\vec{k})\alpha (\vec{k}, -1) \right]
\exp \{i(\vec{k} \cdot \vec{x}-\omega_k t) \} 
\ee
where ${\bf e}(\vec{k}), {\bf e}^{\ast}(\vec{k})$ satisfy \eqref{264b}, \eqref{269b} and $\alpha (\vec{k}, +1), \alpha (\vec{k}, -1)$  are arbitrary scalar functions of $\vec{k}.$
Factor $\sqrt{\hbar c} $ has been taken for further convenience.

Formula \eqref{34} can be rewritten in a form
\be
\label{35}
{\bf \Psi}(\vec{x},t)={\bf \Psi}_+(\vec{x},t)+ {\bf \Psi}_-(\vec{x},t),
\ee
where functions 
\[
{\bf \Psi}_+(\vec{x},t)= \sqrt{\hbar c} 
\int_{{\mathbb R}^3}\frac{d^3 k}{(2 \pi)^3} \,  {\bf e}(\vec{k})
\alpha (\vec{k}, +1)
\exp \{i(\vec{k} \cdot \vec{x}-\omega_k t) \}, 
\]
\be
\label{36}
{\bf \Psi}_-(\vec{x},t)= \sqrt{\hbar c} 
\int_{{\mathbb R}^3}\frac{d^3 k}{(2 \pi)^3} \,  {\bf e}^{\ast}(\vec{k})
\alpha (\vec{k}, -1)
\exp \{i(\vec{k} \cdot \vec{x}-\omega_k t) \}
\ee
are superpositions of $(+1)$ -- helicity states and $(-1)$ -- helicity states respectively. 

One easily finds that ${\bf \Psi}_+(\vec{x},t)$ and ${\bf \Psi}_-(\vec{x},t)$ fulfill the following equations
\begin{subequations}
\label{37}
\begin{align}
i \hbar \,\partial_t {\bf \Psi}_+ &= c \left(\vec{\mathcal S} \cdot \widehat{\vec{p}} \,\right) {\bf \Psi}_+\, , 
\label{37a} \\
i \hbar \,\partial_t {\bf \Psi}_- &= - c \left(\vec{\mathcal S} \cdot \widehat{\vec{p}} \,\right) {\bf \Psi}_-\,.
\label{37b}
\end{align}
\end{subequations}
Let $\widehat{\Pi}_+$ and $\widehat{\Pi}_-$ denote projection operators
\be
\label{38}
\widehat{\Pi}_+ {\bf \Psi} ={\bf \Psi}_+ \quad, \quad \widehat{\Pi}_- {\bf \Psi} = {\bf \Psi}_-\; .
\ee
By adding Eqs. \eqref{37a} and \eqref{37b} we can see that the photon wave function satisfies the Schroedinger -- like evolution equation  
\be
\label{39}
i \hbar \,\partial_t {\bf \Psi} = \widehat{H} {\bf \Psi}
\ee
where $\widehat{H}$ is the Hamilton operator
\be
\label{310}
\widehat{H} :=  c \left(\vec{\mathcal S} \cdot \widehat{\vec{p}} \,\right) \left(\widehat{\Pi}_+ - \widehat{\Pi}_- \right).
\ee
${\bf \Psi}$ is a $1$ -- column matrix
\be
\label{311}
{\bf \Psi}(\vec{x},t) = \left(
\begin{array}{c} \Psi_1(\vec{x},t) \\
\Psi_2(\vec{x},t) \\
\Psi_3(\vec{x},t)
\end{array} \right) 
\ee
and by \eqref{264b} and \eqref{269b} is subject to the constraint equation
\be
\label{312}
\sum_{j=1}^3 \partial_j \Psi_j=0.
\ee
Therefore the photon wave function components satisfy the following systems of differential equations
\setlength{\extrarowheight}{4pt}
\begin{subequations}
\label{37Again}
\begin{align}
\left(\begin{array}{c}
\frac{\partial \Psi_{+1}}{\partial t} \\
\frac{\partial \Psi_{+2}}{\partial t} \\
\frac{\partial \Psi_{+3}}{\partial t}
\end{array}
\right)= -ic
\left(\begin{array}{c}
- \frac{\partial \Psi_{+2}}{\partial x^3} + \frac{\partial \Psi_{+3}}{\partial x^2} \\
 \frac{\partial \Psi_{+1}}{\partial x^3} - \frac{\partial \Psi_{+3}}{\partial x^1} \\
- \frac{\partial \Psi_{+1}}{\partial x^2} + \frac{\partial \Psi_{+2}}{\partial x^1}
\end{array}
\right) \quad , \quad \frac{\partial \Psi_{+1}}{\partial x^1} + \frac{\partial \Psi_{+2}}{\partial x^2} + \frac{\partial \Psi_{+3}}{\partial x^3}=0, \\
\left(\begin{array}{c}
\frac{\partial \Psi_{-1}}{\partial t} \\
\frac{\partial \Psi_{-2}}{\partial t} \\
\frac{\partial \Psi_{-3}}{\partial t}
\end{array}
\right)= ic
\left(\begin{array}{c}
- \frac{\partial \Psi_{-2}}{\partial x^3} + \frac{\partial \Psi_{-3}}{\partial x^2} \\
 \frac{\partial \Psi_{-1}}{\partial x^3} - \frac{\partial \Psi_{-3}}{\partial x^1} \\
- \frac{\partial \Psi_{-1}}{\partial x^2} + \frac{\partial \Psi_{-2}}{\partial x^1}
\end{array}
\right) \quad , \quad \frac{\partial \Psi_{-1}}{\partial x^1} + \frac{\partial \Psi_{-2}}{\partial x^2} + \frac{\partial \Psi_{-3}}{\partial x^3}=0.
\end{align}
\end{subequations}
\setlength{\extrarowheight}{0pt}
A natural question arises whether there exists any probabilistic interpretation of the photon wave function ${\bf \Psi}(\vec{x},t).$ In the Dirac notation one can write ${\bf \Psi}(\vec{x},t)=\big<\vec{x} \big|{\bf  \Psi}(t) \big>$ with
\setlength{\extrarowheight}{4pt}
\be
\label{313}
 \big| {\bf \Psi}(t) \big>= \left(\begin{array}{c}
 \big| \Psi_1(t) \big>\\
 \big| \Psi_2(t) \big>\\
 \big| \Psi_3(t) \big>
 \end{array}
 \right).
\ee
\setlength{\extrarowheight}{0pt}
Therefore ${\bf \Psi}(\vec{x},t)$ represents a state of photon in position representation. However, the Hermitian operator $\widehat{\vec{x}}$ cannot be considered as a photon position operator or any other photon observable since $\vec{x}\,\Psi(\vec{x},t) $ does not satisfy the constraint condition \eqref{312} although ${\bf \Psi}(\vec{x},t)$ does. Consequently, quantity 
\[
\int_{V} d^3 x \,  {\bf \Psi}^{\dagger}(\vec{x},t) {\bf \Psi}(\vec{x},t)
\]
 cannot be interpreted as the probability of finding the photon in the domain $V$ at instant $t$ like in ``usual'' quantum mechanics. 

A correct interpretation seems to be the one given by I. Bia\l ynicki -- Birula \cite{Bialynicki1994, Bialynicki1996} and J. E. Sipe \cite{Sipe1995}. Assume that ${\bf \Psi}(\vec{x},t)$ is normalised as 
\be
\label{314}
\int_{{\mathbb R}^3} d^3 x \,  {\bf \Psi}^{\dagger}(\vec{x},t) {\bf \Psi}(\vec{x},t)= \big< E \big>(t),
\ee
 where $\big< E \big>(t) $ denotes the average energy of photon at instant $t$. Condition \eqref{314} is a counterpart of classical formula \eqref{274a}. Inserting \eqref{34} into \eqref{314} one receives
 \[
 \big< E \big>(t)= \int_{{\mathbb R}^3} \frac{d^3 k}{(2 \pi)^3 |\vec{k}|}\, \hbar \omega_k
 \left(|\alpha(\vec{k},+1)|^2 + |\alpha(\vec{k},-1)|^2
 \right)
 \]
 \be
 \label{315}
 = \int_{{\mathbb R}^3} \frac{d^3 k}{(2 \pi)^3 |\vec{k}|}\, \hbar \omega_k 
  \widetilde{\bf \Psi}^{\dagger}(\vec{k}) \widetilde{\bf \Psi}(\vec{k}),
 \ee
 where 
 \[
 \widetilde{\bf \Psi}(\vec{k}):= \frac{1}{\sqrt{\hbar c}} \int_{{\mathbb R}^3} d^3x \,{\bf \Psi}(\vec{x},0) \exp \left\{ - i \vec{k} \cdot \vec{x} \right\}
 \]
 \be
 \label{316}
 = {\mathbf e}(\vec{k}) \alpha(\vec{k},+1) + {\mathbf e}^{\ast}(\vec{k}) \alpha(\vec{k},-1).
 \ee
 Therefore we see that $\frac{d \left< E \right>(t)}{dt}=0.$ Thus  normalisation \eqref{314} is time independent as expected.
 
 From \eqref{34} with \eqref{316} we quickly find that
 \be
 \label{317}
 \widetilde{\bf \Psi}(\vec{k},t) := \widetilde{\bf \Psi}(\vec{k}) 
 \exp \left\{ - i \omega_k t \right\}=
 \frac{1}{\sqrt{\hbar c}} \int_{{\mathbb R}^3} d^3x \,{\bf \Psi}(\vec{x},t) \exp \left\{ - i \vec{k} \cdot \vec{x} \right\}.
 \ee
 Formula \eqref{315}  justifies the conclusion that the integral
 \be
 \label{318}
\wp \left(\vec{k} \in \Omega, \lambda\right):= \int_{\Omega} \frac{d^3 k}{(2 \pi)^3 |\vec{k}|} \left| \alpha(\vec{k}, \lambda)\right|^2 \quad , \quad \lambda= \pm 1
 \ee
 represents the \textit{probability that $\vec{k}=(k_1, k_2,k_3) \in \Omega \subset {\mathbb R}^3$ and the helicity of photon is $\lambda$.} This probability does not depend on time.
 
 Then the integral
 \be
 \label{319}
 \wp \left(\vec{k} \in \Omega \right) := \int_{\Omega} \frac{d^3 k}{(2 \pi)^3 |\vec{k}|}
  \widetilde{\bf \Psi}^{\dagger}(\vec{k},t) \widetilde{\bf \Psi}(\vec{k},t)=
  \int_{\Omega} \frac{d^3 k}{(2 \pi)^3 |\vec{k}|}
  \widetilde{\bf \Psi}^{\dagger}(\vec{k}) \widetilde{\bf \Psi}(\vec{k})
 \ee
 is the \textit{probability that $\vec{k}=(k_1, k_2,k_3) \in \Omega \subset {\mathbb R}^3$.}
 Probability $ \wp \left(\vec{k} \in \Omega \right)$ is independent of time.
 
 If one assumes that these conclusions are true then the average momentum of photon is given by
 \[
  \big< \vec{p} \,\big>(t) = \big< \hbar \vec{k} \big>=
 \int_{{\mathbb R}^3} \frac{d^3 k}{(2 \pi)^3 |\vec{k}|}\, \hbar \vec{k}\,
 \widetilde{\bf \Psi}^{\dagger}(\vec{k},t) \widetilde{\bf \Psi}(\vec{k},t)
 \]
 \be
 \label{320}
 =\int_{{\mathbb R}^3} \frac{d^3 k}{(2 \pi)^3 |\vec{k}|}\, \hbar \vec{k}\,
 \left(|\alpha(\vec{k},+1)|^2 + |\alpha(\vec{k},-1)|^2
 \right).
 \ee
 As in the case of average energy, the mean value of momentum remains constant.
 Formula \eqref{320} suits perfectly the classical relation \eqref{274b}. A slight difference is caused by additional factor $\sqrt{\hbar c}$ in ${\bf \Psi}(\vec{x},t)$ (see \eqref{34}).
 
 Moreover,
 \be
 \label{321}
 \big< E_V \big>(t):= \int_{V} d^3 x \, {\bf \Psi}^{\dagger}(\vec{x},t) {\bf \Psi}(\vec{x},t)
 \ee
 is the \textit{average (expected) energy of photon detected in volume $V \subset {\mathbb R}^3$} at instant $t$. Thus
 \be
 \label{322}
 \wp \left(\vec{x} \in V,t \right):=\frac{\int_{V} d^3 x \, {\bf \Psi}^{\dagger}(\vec{x},t) {\bf \Psi}(\vec{x},t)}{\int_{{\mathbb R}^3} d^3 x \, {\bf \Psi}^{\dagger}(\vec{x},t) {\bf \Psi}(\vec{x},t)}
 \ee
 determines the \textit{probability that the photon energy is localised in volume $V \subset {\mathbb R}^3$ at instant $t$.}
 
 Note that this interpretation of formula \eqref{322} remains true also when normalisation of 
 ${\bf \Psi}(\vec{x},t)$ is different than proposed in \eqref{314}. Then formulae \eqref{318} and \eqref{319} turn into
 \be
 \label{323}
 \wp \left(\vec{k} \in \Omega, \lambda\right)= \frac{\int_{\Omega} \frac{d^3 k}{(2 \pi)^3 |\vec{k}|} \left| \alpha(\vec{k}, \lambda)\right|^2}
 {\int_{{\mathbb R}^3}  \frac{d^3 k}{(2 \pi)^3 |\vec{k}|} \, \widetilde{\bf \Psi}^{\dagger}(\vec{k}) \widetilde{\bf \Psi}(\vec{k})}
 \ee
 and
 \be
 \label{324}
 \wp \left(\vec{k} \in \Omega \right)=\frac
  {\int_{\Omega}  \frac{d^3 k}{(2 \pi)^3 |\vec{k}|} \, \widetilde{\bf \Psi}^{\dagger}(\vec{k}) \widetilde{\bf \Psi}(\vec{k})}
   {\int_{{\mathbb R}^3}  \frac{d^3 k}{(2 \pi)^3 |\vec{k}|} \, \widetilde{\bf \Psi}^{\dagger}(\vec{k}) \widetilde{\bf \Psi}(\vec{k})}
 \ee
 respectively.

In Sec. \ref{sec2} the electromagnetic field was represented by the Riemann -- Silberstein matrix ${\mathbf F}$ and equivalently by matrix ${\mathbf f}.$ These two representations are related by a unitary transformations \eqref{244}. In quantum model of photon a  wave function ${\bf \Psi}\left( \vec{x},t \right)$ based on the   Riemann -- Silberstein matrix has been built.

 Let us consider now a photon wave function obtained from wave function ${\bf \Psi}\left( \vec{x},t \right)$ by a unitary transformation defined in classical electrodynamics by \eqref{244} with \eqref{245}. So one puts
 \be
 \label{325}
 {\bf \Psi}'\left( \vec{x},t \right)= {\mathcal U} \,{\bf \Psi}\left( \vec{x},t \right).
 \ee
 Inserting \eqref{34} into \eqref{325} and applying \eqref{245} we receive
 \[
 {\bf \Psi}'\left( \vec{x},t \right)= \sqrt{\hbar c}\int_{{\mathbb R}^3}\frac{d^3 k}{(2 \pi)^3}
 \left[
 {\mathcal U} \cdot {\bf e}(\vec{k})
\alpha (\vec{k}, +1)
+ {\mathcal U} \cdot {\mathcal U}^{\rm T} \left({\mathcal U} \cdot {\bf e}(\vec{k}) \right)^{\ast}\alpha (\vec{k}, -1)
 \right] \exp \{i(\vec{k} \cdot \vec{x}-\omega_k t) \}
 \]
 \setlength{\extrarowheight}{2pt}
 \[
 = \sqrt{\hbar c}\int_{{\mathbb R}^3}\frac{d^3 k}{(2 \pi)^3}
 \left[
 \left(\begin{array}{r}
 e_{\accentset{\mbox{\huge .} }{1} \accentset{\mbox{\huge .} }{1}}(\vec{k}) \\
 \sqrt{2}  e_{\accentset{\mbox{\huge .} }{1} \accentset{\mbox{\huge .} }{2}}(\vec{k}) \\
  e_{\accentset{\mbox{\huge .} }{2} \accentset{\mbox{\huge .} }{2}}(\vec{k})
\end{array} \right) \alpha (\vec{k}, +1) -
\left(\begin{array}{r}
 e_{\accentset{\mbox{\huge .} }{2} \accentset{\mbox{\huge .} }{2}}(\vec{k}) \\
 -\sqrt{2}  e_{\accentset{\mbox{\huge .} }{1} \accentset{\mbox{\huge .} }{2}}(\vec{k}) \\
  e_{\accentset{\mbox{\huge .} }{1} \accentset{\mbox{\huge .} }{1}}(\vec{k})
\end{array} \right)^{\ast} \alpha (\vec{k}, -1)
 \right] \exp \{i(\vec{k} \cdot \vec{x}-\omega_k t) \}
 \]
 \be
 \label{326}
 = \sqrt{\hbar c}\int_{{\mathbb R}^3}\frac{d^3 k}{(2 \pi)^3}
 \left[
 \left(\begin{array}{r}
 e_{\accentset{\mbox{\huge .} }{1} \accentset{\mbox{\huge .} }{1}}(\vec{k}) \\
 \sqrt{2}  e_{\accentset{\mbox{\huge .} }{1} \accentset{\mbox{\huge .} }{2}}(\vec{k}) \\
  e_{\accentset{\mbox{\huge .} }{2} \accentset{\mbox{\huge .} }{2}}(\vec{k})
\end{array} \right) \alpha (\vec{k}, +1) -
\left(\begin{array}{r}
 e^{11}(\vec{k}) \\
 \sqrt{2}  e^{12}(\vec{k}) \\
  e^{22}(\vec{k})
\end{array} \right) \alpha (\vec{k}, -1)
\right] \exp \{i(\vec{k} \cdot \vec{x}-\omega_k t) \}
 \ee
 where $ e_{\accentset{\mbox{\huge .} }{A} \accentset{\mbox{\huge .} }{B}}(\vec{k})$ and
 $e^{AB}(\vec{k}) $ are symmetric spinors defined as
 \be
 \label{327}
 e_{\accentset{\mbox{\huge .} }{A} \accentset{\mbox{\huge .} }{B}}(\vec{k}):= i\, \Phi^j_{\, \accentset{\mbox{\huge .} }{A} \accentset{\mbox{\huge .} }{B}} e_j(\vec{k}) \quad , \quad 
 e^{AB}(\vec{k}):= \left(i\, \Phi^{j \, \accentset{\mbox{\huge .} }{A} \accentset{\mbox{\huge .} }{B}} e_j(\vec{k})  \right)^{\ast}
 \ee
 (see \eqref{220}, \eqref{221} and \eqref{223}).
 
 Photon wave function ${\bf \Psi}'\left( \vec{x},t \right)$ satisfies the evolution Schroedinger -- like equation
 \be
 \label{328}
 i \hbar \partial_t {\bf \Psi}' = \widehat{H}' {\bf \Psi}',
 \ee
 where the Hamilton operator
 \be
 \label{329}
 \widehat{H}' = c \left( \vec{\mathcal S}' \cdot \widehat{\vec{p}}\right) \cdot
 \left(\widehat{\Pi}'_{+} - \widehat{\Pi}'_{-} \right),
 \ee
 \[
 \widehat{\Pi}'_{\pm} = {\mathcal U}\,\widehat{\Pi}_{\pm}\,{\mathcal U}^{\dagger}
 \]
 and with $\vec{\mathcal S}' $ given by \eqref{241}. Comparing \eqref{326} with \eqref{35} and \eqref{36} one concludes that ${\bf \Psi}'\left( \vec{x},t \right)$ can be expressed as
 \be
 \label{330}
 {\bf \Psi}'\left( \vec{x},t \right)= {\bf \Psi}'_+\left( \vec{x},t \right) + {\bf \Psi}'_-\left( \vec{x},t \right),
 \ee
 where 
 \be
 \label{331}
 {\bf \Psi}'_+\left( \vec{x},t \right)= 
 \sqrt{\hbar c}\int_{{\mathbb R}^3}\frac{d^3 k}{(2 \pi)^3}
 \left(\begin{array}{r}
 e_{\accentset{\mbox{\huge .} }{1} \accentset{\mbox{\huge .} }{1}}(\vec{k}) \\
 \sqrt{2}  e_{\accentset{\mbox{\huge .} }{1} \accentset{\mbox{\huge .} }{2}}(\vec{k}) \\
  e_{\accentset{\mbox{\huge .} }{2} \accentset{\mbox{\huge .} }{2}}(\vec{k})
\end{array} \right) \alpha (\vec{k}, +1) \exp \{i(\vec{k} \cdot \vec{x}-\omega_k t) \}
=\widehat{\Pi}'_{+} {\bf \Psi}' \left( \vec{x},t \right)
 \ee
 consists of states of helicity $+1$
 \be
 \label{332}
 \frac{\vec{\mathcal S}' \cdot \widehat{\vec{p}}}{|\vec{p}\,|} {\bf \Psi}'_+\left( \vec{x},t \right)=
 +1 \cdot {\bf  \Psi}'_+\left( \vec{x},t \right)
 \ee
 and 
 \be
 \label{333}
 {\bf  \Psi}'_-\left( \vec{x},t \right)=  \sqrt{\hbar c}\int_{{\mathbb R}^3}\frac{d^3 k}{(2 \pi)^3}
  \left(\begin{array}{r}
 e^{11}(\vec{k}) \\
 \sqrt{2}  e^{12}(\vec{k}) \\
  e^{22}(\vec{k})
\end{array} \right) \left( -\alpha (\vec{k}, -1) \right)
\exp \{i(\vec{k} \cdot \vec{x}-\omega_k t) \}= \widehat{\Pi}'_{-} {\bf \Psi}' \left( \vec{x},t \right)
 \ee
is a superposition of states of helicity $-1$ 
\be
\label{334}
\frac{\vec{\mathcal S}' \cdot \widehat{\vec{p}}}{|\vec{p}\,|} {\bf \Psi}'_-\left( \vec{x},t \right)=
 -1 \cdot  {\bf \Psi}'_-\left( \vec{x},t \right).
\ee
 We end this section with a formula giving relation between $\big< \vec{k} \big|{\bf \Psi}(t) \big>$ and $\widetilde{\bf \Psi} (\vec{k},t).$ We have
 \[
  \big< \vec{k} \big| {\bf \Psi}(t) \big>= \int_{{\mathbb R}^3} d^3 x \big< \vec{k} \big| \vec{x} \big> \big< \vec{x} \big| {\bf \Psi}(t) \big> = 
 \frac{1}{(2\pi)^{3/2}} \int_{{\mathbb R}^3} d^3 x \, {\bf \Psi}(\vec{x},t) \exp \left\{- i \vec{k} \cdot \vec{x} \right\}
 \]
 \be
 \label{335}
 =\sqrt{\frac{\hbar c}{(2 \pi)^3}} \widetilde{\bf \Psi} (\vec{k},t),
 \ee
 where Eq. \eqref{317} has been used.
 
 
 \section{Scalar product, generalized Hermitian operators and observables. The density operator.}
\label{sec4}
\setcounter{equation}{0}

One quickly sees that the formula (\ref{319}) leads to the obvious normalization of ${\bf \Psi}(\vec{x},t)$
\begin{equation}\label{41}
  \int_{{\mathbb R}^3} \frac{d^{3}k}{(2\pi)^{3}|\vec{k}|}\widetilde{\bf \Psi}^{\dagger}(\vec{k})\widetilde{\bf \Psi}(\vec{k})=1 \quad \Longleftrightarrow \quad \int_{{\mathbb R}^3} d^{3}x {\bf \Psi}^{\dagger}(\vec{x},t)\widehat{H}^{-1}{\bf \Psi}(\vec{x},t)=1.
\end{equation}
In the Dirac notation we write
\begin{equation}\label{42}
  \langle{\bf \Psi}(t)|\widehat{H}^{-1}|{\bf \Psi}(t)\rangle =1.
\end{equation}
 Due to the facts that energy and momentum  are constants of motion, formulae \eqref{314} or \eqref{315} and \eqref{320} take forms
\begin{equation}\label{43}
  \langle E \rangle= \langle {\bf \Psi}(t)|\widehat{H}^{-1}\widehat{H}|{\bf \Psi}(t)\rangle
\end{equation}
and
\begin{equation}\label{44}
  \langle \vec{p} \,\rangle =\langle {\bf \Psi}(t)|\widehat{H}^{-1}\,\widehat{\vec{p}} \,|{\bf \Psi}(t)\rangle
\end{equation}
respectively. These last two formulae suggest that it is reasonable to introduce a new scalar product which will be called the \textit{Bia{\l}ynicki -- Birula scalar product} \cite{Bialynicki1996, Hawton1999a}
\begin{equation}\label{45}
   \langle {\bf \Psi}_{1}|{\bf \Psi}_{2}\rangle_{BB} := \langle{\bf \Psi}_{1}|\widehat{H}^{-1}|{\bf \Psi}_{2}\rangle.
\end{equation}
Straightforward calculations show that \cite{Bialynicki1996}
\[
   \langle {\bf \Psi}_{1}|{\bf \Psi}_{2}\rangle_{BB} = \int_{{\mathbb R}^3}\frac{d^{3}k}{(2\pi)^{3}|\vec{k}|}\widetilde{\bf \Psi}_{1}^{\dagger}(\vec{k})\widetilde{\bf \Psi}_{2}(\vec{k}) 
   = \sum_{\lambda=-1,1}\int_{{\mathbb R}^3} \frac{d^{3}k}{(2\pi)^{3}|\vec{k}|}\alpha_{1}^{\ast}(\vec{k},\lambda)\alpha_{2}(\vec{k},\lambda) 
  \] 
   \be
   \label{46}
   = \frac{1}{2\pi^{2}\hbar c} \int_{{\mathbb R}^6} d^{3}x d^{3}x' {\bf \Psi}_{1}^{\dagger}(\vec{x})\frac{1}{|\vec{x}-\vec{x}'|^{2}}{\bf \Psi}_{2}(\vec{x'}).
\ee
If $\widehat{\gamma}$ is a linear operator then we define the \textit{generalized Hermitian conjugation of} $\widehat{\gamma}$ as a linear operator $\widehat{\gamma}^{+\hspace{-0.45em}+}$ such that
\begin{equation}\label{47}
  \langle{\bf \Psi}_{1}|\widehat{\gamma}|{\bf \Psi}_{2}\rangle_{BB}^{\ast} = \langle{\bf \Psi}_{2}|\widehat{\gamma}^{+\hspace{-0.45em}+}|{\bf \Psi}_{1}\rangle_{BB}
\end{equation}
[compare with \cite{Feshbach1958, Davydov1976}].

From (\ref{45}) one gets
\begin{align}\label{48}
\langle{\bf \Psi}_{1}|\widehat{\gamma}|{\bf \Psi}_{2}\rangle_{BB}^{\ast} &= \langle{\bf \Psi}_{1}|\widehat{H}^{-1}\widehat{\gamma}|{\bf \Psi}_{2}\rangle^{\ast}=\langle{\bf \Psi}_{2}|\widehat{\gamma}^{\dag}\widehat{H}^{-1}|{\bf \Psi}_{1}\rangle \nonumber \\
&=\langle{\bf \Psi}_{2}|\widehat{H}^{-1}\widehat{H}\widehat{\gamma}^{\dag}\widehat{H}^{-1}|{\bf \Psi}_{1}\rangle=
\langle{\bf \Psi}_{2}|\widehat{H}\widehat{\gamma}^{\dag}\widehat{H}^{-1}|{\bf \Psi}_{1}\rangle_{BB}.
\end{align}
Comparing (\ref{48}) with (\ref{47}) we find
\begin{equation}\label{49}
  \widehat{\gamma}^{\,+\hspace{-0.45em}+} = \widehat{H}\, \widehat{\gamma}^{\dag}\, \widehat{H}^{-1}.
\end{equation}
Hence
\begin{equation}\label{410}
 \langle{\bf \Psi}|\widehat{\gamma}|{\bf \Psi}\rangle_{BB}^{\ast}=\langle{\bf \Psi}|\widehat{\gamma}|{\bf \Psi}\rangle_{BB}
\end{equation}
for any $|{\bf \Psi}\rangle$ if and only if
\begin{equation}\label{411}
  \widehat{\gamma}^{\,+\hspace{-0.45em}+}=\widehat{\gamma} \quad \Longleftrightarrow \quad \widehat{\gamma}^{\dag}=\widehat{H}^{-1}\widehat{\gamma}\widehat{H}.
\end{equation}
 Condition (\ref{411}) is called the \textit{generalized hermicity condition} and a linear operator satisfying (\ref{411}) will be called the \textit{generalized Hermitian operator}. One easily shows that
\begin{equation}\label{412}
  \widehat{\vec{p}}^{\,+\hspace{-0.45em}+}=\widehat{\vec{p}}, \quad \widehat{H}^{+\hspace{-0.45em}+}=\widehat{H}, \quad \widehat{\vec{x}}^{\,+\hspace{-0.45em}+}\neq\widehat{\vec{x}}.
\end{equation}
From (\ref{43}), (\ref{44}) and (\ref{45}) we conclude that if $\widehat{\vartheta}$ is a photon observable and $|{\bf \Psi}\rangle$ is the photon state normalized according to (\ref{42})
$\langle{\bf \Psi}|{\bf \Psi}\rangle_{BB}=1$ then the average (expected) value of this observable in  state $|{\bf \Psi}\rangle$ reads
\begin{equation}\label{413}
   \langle \widehat{\vartheta}\rangle=\langle{\bf \Psi}|\widehat{\vartheta}|{\bf \Psi}\rangle_{BB}=\langle{\bf \Psi}|\widehat{H}^{-1}\widehat{\vartheta}|{\bf \Psi}\rangle.
\end{equation}
Since $\langle \widehat{\vartheta}\rangle$ must be real for arbitrary $|{\bf \Psi}\rangle$  operator $\widehat{\vartheta}$ has to be a generalized Hermitian operator. So if $\widehat{\vartheta}$ is any photon observable then necessary
\begin{equation}\label{414}
  \widehat{\vartheta}^{+\hspace{-0.45em}+}=\widehat{\vartheta} \quad \Longleftrightarrow \quad \widehat{\vartheta}^{\dag}=\widehat{H}^{-1}\widehat{\vartheta}\widehat{H}.
\end{equation}
Obviously, in the general case, when $|{\bf \Psi}\rangle$ is not normalized as above, one has
\begin{equation}\label{415}
  \langle \widehat{\vartheta}\rangle=\frac{\langle{\bf \Psi}|\widehat{\vartheta}|{\bf \Psi}\rangle_{BB}}{\langle{\bf \Psi}|{\bf \Psi}\rangle_{BB}}
  = \frac{\langle{\bf \Psi}|\widehat{H}^{-1}\widehat{\vartheta}|{\bf \Psi}\rangle}{\langle{\bf \Psi}|\widehat{H}^{-1}|{\bf \Psi}\rangle}.
\end{equation}
We rewrite (\ref{415}) in the form
\begin{equation}\label{416}
  \langle \widehat{\vartheta}\rangle=\frac{\langle{\bf \Psi}|\widehat{H}^{-1/2}\widehat{H}^{-1/2}\,\widehat{\vartheta}\,
  \widehat{H}^{1/2}\widehat{H}^{-1/2}|{\bf \Psi}\rangle}{\langle{\bf \Psi}|\widehat{H}^{-1/2}\widehat{H}^{-1/2}|{\bf \Psi}\rangle}.
\end{equation}
Hence one infers that the \textit{average value of $\widehat{\vartheta}$ in the state $|{\bf \Psi}\rangle$ is equal to the average value of the operator}
\begin{equation}\label{417}
  \widehat{\vartheta}_{H}:=\widehat{H}^{-1/2}\,\widehat{\vartheta}\,\widehat{H}^{1/2}
\end{equation}
\textit{in the state $\widehat{H}^{-1/2}|{\bf \Psi}\rangle$, calculated in accordance with the usual scalar product $\langle \cdot | \cdot\rangle$}.

From (\ref{414}) and (\ref{417}) we find that
\begin{equation}\label{418}
  \widehat{\vartheta}^{+\hspace{-0.45em}+}=\widehat{\vartheta} \quad \Longleftrightarrow \quad \widehat{\vartheta}_{H}^{\,\dag}=\widehat{\vartheta}_{H}.
\end{equation}
If $| \pmb{\vartheta}\rangle$ is an eigenvector of a photon observable $\widehat{\vartheta}$ corresponding to the eigenvalue $\vartheta$
\begin{equation}\label{419}
 \widehat{\vartheta}|{\pmb \vartheta}\rangle=\vartheta|{\pmb \vartheta}\rangle, \quad \widehat{\vartheta}^{+\hspace{-0.45em}+}=\widehat{\vartheta}
\end{equation}
then $\widehat{H}^{-1/2}|\pmb{\vartheta}\rangle$ is an eigenvector of the Hermitian operator $\widehat{\vartheta}_{H}$ given by (\ref{417}) corresponding to the same eigenvalue $\vartheta$
\begin{equation}\label{420}
  \widehat{\vartheta}_{H}\widehat{H}^{-1/2}|\pmb{\vartheta}\rangle=\vartheta\widehat{H}^{-1/2}|\pmb{\vartheta}\rangle.
\end{equation}
Therefore, keeping also in mind  formula (\ref{416}), we can conclude that \textit{having given a photon in  state $|{\bf \Psi}\rangle$ one can equivalently consider it as a quantum particle in  state $\widehat{H}^{-1/2}|{\bf \Psi}\rangle$, with observables defined by  transformation 
\[
\widehat{\vartheta} \longmapsto \widehat{\vartheta}_{H}=\widehat{H}^{-1/2}\widehat{\vartheta}\widehat{H}^{1/2}.
\]
 However the scalar product used now is the usual scalar product $\langle \cdot | \cdot\rangle$}.
 
 Let $|{\bf \Psi}\rangle$ be a photon state vector normalized to $1$ with respect to the Bia{\l}ynicki-Birula scalar product (\ref{45}), $\langle {\bf \Psi}|{\bf \Psi}\rangle_{BB}=1$, and let $\widehat{\vartheta}$ be a photon observable satisfying the generalized hermicity condition (\ref{414}). Then by (\ref{415}) the average value of $\widehat{\vartheta}$ in  state $|{\bf \Psi}\rangle$ can be written as
\be
\label{421}
  \langle \widehat{\vartheta}\rangle=\langle{\bf \Psi}|\widehat{H}^{-1}\widehat{\vartheta}|{\bf \Psi}\rangle 
  = \textrm{Tr}\left\{\widehat{H}^{-1/2}\widehat{\vartheta}\widehat{H}^{1/2}\widehat{H}^{-1/2}|{\bf \Psi}\rangle \langle{\bf \Psi}|\widehat{H}^{-1/2}\right\} 
 =  \textrm{Tr}\left\{\widehat{\vartheta}|{\bf \Psi}\rangle \langle{\bf \Psi}|\widehat{H}^{-1}\right\}.
\ee
Motivated by the respective formula for the average value of any observable in nonrelativistic quantum mechanics we define the \textit{density operator for a photon pure state} $|{\bf \Psi}\rangle$, $\langle{\bf \Psi}|{\bf \Psi}\rangle_{BB}=1$, as
\begin{equation}\label{422}
  \widehat{\rho}=|{\bf \Psi}\rangle \langle{\bf \Psi}|\widehat{H}^{-1}.
\end{equation}
This operator fulfills the following properties
\begin{align}\label{423}
 \textrm{(i)} & \enspace \widehat{\rho}^{\,+\hspace{-0.45em}+}=\widehat{\rho}, \nonumber \\
 \textrm{(ii)} & \enspace \textrm{Tr}\left\{\widehat{\rho}\right\}=1, \nonumber \\
 \textrm{(iii)} & \enspace \langle \pmb{\chi}|\widehat{\rho}|\pmb{\chi}\rangle_{BB}\geq 0 \quad \textrm{for every} \enspace |\pmb{\chi}\rangle,  \nonumber\\
 \textrm{(iv)} & \enspace \widehat{\rho}^{\,2}=\widehat{\rho}. 
\end{align}
By analogy to (\ref{417}) one introduces an operator
\begin{equation}\label{424}
  \widehat{\rho}_{H}=\widehat{H}^{-1/2}\,\widehat{\rho}\,\widehat{H}^{1/2}=\widehat{H}^{-1/2}|{\bf \Psi}\rangle \langle{\bf \Psi}|\widehat{H}^{-1/2}.
\end{equation}
It satisfies
\begin{align}\label{425}
\textrm{(i')} & \enspace \widehat{\rho}_{H}^{\,\dag}=\widehat{\rho}_{H}, \nonumber \\
\textrm{(i'i')} & \enspace \textrm{Tr}\left\{\widehat{\rho}_{H}\right\}=1, \nonumber \\
\textrm{(i'i'i')} & \langle \pmb{\chi}|\widehat{\rho}_{H}|\pmb{\chi}\rangle \geq 0 \quad \textrm{for every} \enspace |\pmb{\chi}\rangle, \nonumber \\
\textrm{(i'v')} & \widehat{\rho}_{H}^{\,2}=\widehat{\rho}_{H}. 
\end{align}
Employing the above results concerning the density operator for a pure state of photon we assume that in the general case of pure or mixed photon state this state is represented by  operator $\widehat{\rho}$ fulfilling the conditions (i), (ii) and (iii) of (\ref{423}). This operator is called the \textit{density operator for the photon state}. Density operator $\widehat{\rho}$ defines uniquely  operator $\widehat{\rho}_{H}$ according to the first equality of (\ref{424}). Operator $\widehat{\rho}_{H}$ satisfies the conditions (i'), (i'i') and (i'i'i') of (\ref{425}).

The average (expected) value of any observable $\widehat{\vartheta}$ reads
\begin{equation}\label{426}
  \langle \widehat{\vartheta}\rangle=\textrm{Tr}\left\{\widehat{\vartheta}\widehat{\rho}\right\}=\textrm{Tr}\left\{\widehat{\vartheta}_{H}\widehat{\rho}_{H}\right\}
\end{equation}
where $\widehat{\vartheta}_{H}$ is given by (\ref{417}).

Finally, a photon state is pure iff $\widehat{\rho}$ fulfills the condition (iv) of (\ref{423}) or, equivalently, $\widehat{\rho}_{H}$ fulfills (i'v') of (\ref{425}). The density operator satisfies the Liouville -- von Neumann evolution equation
\begin{equation}\label{427}
  i \hbar\frac{\partial \widehat{\rho}}{\partial t}=\left[\widehat{H}, \widehat{\rho}\right].
\end{equation}
Analogously
\begin{equation}\label{428}
  i \hbar\frac{\partial \widehat{\rho}_{H}}{\partial t}=\left[\widehat{H}, \widehat{\rho}_{H}\right]
\end{equation}
[Remark: It is obvious that we can easily write down all the formulae in the representation defined by the $\mathcal{U}$ -- transformation (see (\ref{325})].

\section{The Weyl -- Wigner -- Moyal formalism and the Wigner function for photon}
\label{sec5}
\setcounter{equation}{0}

Now we have at our disposal all elements required  to develop the Weyl-Wigner-Moyal formalism for photon. Our aim is to develop this formalism in close analogy to that considered in our previous works \cite{Przanowski2019, Przanowski2017}.

First we construct the photon phase space. One starts with the Hilbert space
\begin{equation}\label{51}
  \mathcal{H}=L^{2}(\mathbb{R}^{3})\otimes\mathbb{C}^{3}.
\end{equation}
As it has been shown in \cite{Przanowski2019, Przanowski2017}, the associated phase space is
\begin{equation}\label{52}
  \Gamma=\left\{(\vec{p}, \vec{x}, \phi_{m}, n)\right\}= \mathbb{R}^{3}\times \mathbb{R}^{3}\times \Gamma^{3},
\end{equation}
where $\Gamma^{3}$ is a $3\times3$ grid, $\Gamma^{3}=\left\{(\phi_{m}, n)\right\}$ $m, n=0,1,2$, $\phi_{m}=\frac{2\pi}{3}m$.

(In the current paper the position vector is denoted by $\vec{x}$ and not by $q$ as in \cite{Przanowski2019}).

As we remember, a function on a phase space associated to an observable is real. However, the observables considered in the previous cases were  Hermitian operators. In contrary, the photon observables analysed in Section \ref{sec4} are  generalized Hermitian operators. So now one should define a correspondence between operators in $\mathcal{H}$ and functions in $\Gamma$ in such a way that the functions corresponding to the generalized Hermitian operators are real. To this end we proceed as follows. Given unitary operators (see Eqs. (2.9), (2.12), (2.13) and (2.15) in our previous paper \cite{Przanowski2019} for $s+1=3$)
\begin{align}\label{53}
   \widehat{\mathcal{D}}(k,l)&=\exp\left\{i\frac{\pi k l}{3}\right\}\sum_{m=0}^{2}\exp\left\{i\frac{2\pi k m}{3}\right\}|\phi_{m+l}\rangle\langle\phi_{m}| \nonumber \\
   &=\exp\left\{i\frac{\pi k l}{3}\right\}\sum_{n=0}^{2}\exp\left\{i\frac{2\pi n l}{3}\right\}|n\rangle\langle n+k \; mod \, 3|
\end{align}
for $k,l=0,1,2$ and
\[
 \widehat{\mathcal{U}}(\vec{\lambda},\vec{\mu})=\exp\{i(\vec{\lambda}\cdot\widehat{\vec{p}}+\vec{\mu}\cdot\widehat{\vec{x}})\} 
 \]
 \[ 
  =\int\limits_{\mathbb{R}^{3}}d^{3}x \exp\{i\vec{\mu}\cdot \vec{x}\}\left|\vec{x}-\frac{\hbar\vec{\lambda}}{2}\right\rangle \left\langle \vec{x}+\frac{\hbar\vec{\lambda}}{2}\right|
\]
\be
\label{54}
      =\int\limits_{\mathbb{R}^{3}}d^{3}p \exp\{i\vec{\lambda}\cdot \vec{p}\,\}\left|\vec{p}+\frac{\hbar\vec{\mu}}{2}\right\rangle \left\langle \vec{p}-\frac{\hbar\vec{\mu}}{2}\right|
\ee
we define two \textit{generalized unitary operators}
\begin{equation}\label{55}
   \widehat{\widetilde{\mathcal{D}}}(k,l):=\widehat{H}^{1/2}\,\widehat{\mathcal{D}}(k,l)\, \widehat{H}^{-1/2}, \quad
   \widehat{\widetilde{\mathcal{U}}}(\vec{\lambda},\vec{\mu}):=\widehat{H}^{1/2}\,\widehat{\mathcal{U}}(\vec{\lambda},\vec{\mu})\,\widehat{H}^{-1/2}.
\end{equation}
They satisfy the following properties
\begin{subequations}\label{56}
\begin{align}
   \widehat{\widetilde{\mathcal{D}}}^{+ \hspace{-0.45em}+}(k,l)&= \widehat{\widetilde{\mathcal{D}}}^{-1}(k,l)=\widehat{\widetilde{\mathcal{D}}}(-k,-l), \label{56a}\\
   \textrm{Tr}\left\{\widehat{\widetilde{\mathcal{D}}}(k,l)\right\} &= 3\delta_{k0}\delta_{l0}, \quad 0\leq k,l \leq 2, \label{56b} \\
   \textrm{Tr}\left\{\widehat{\widetilde{\mathcal{D}}}(k,l)\widehat{\widetilde{\mathcal{D}}}^{+ \hspace{-0.45em}+}(k',l')\right\} &= 3\delta_{kk'}\delta_{ll'} \quad 0\leq k,l,k',l'\leq 2 \label{56c}
\end{align}
\end{subequations}
and
\begin{subequations}\label{57}
\begin{align}
   \widehat{\widetilde{\mathcal{U}}}^{+ \hspace{-0.45em}+}(\vec{\lambda},\vec{\mu})&= \widehat{\widetilde{\mathcal{U}}}^{-1}(\vec{\lambda},\vec{\mu})=\widehat{\widetilde{\mathcal{U}}}(-\vec{\lambda},-\vec{\mu}), \label{57a}\\
   \textrm{Tr}\left\{\widehat{\widetilde{\mathcal{U}}}(\vec{\lambda},\vec{\mu})\right\} &= \left(\frac{2\pi}{\hbar}\right)^{3}\delta(\vec{\lambda})\delta(\vec{\mu}), \label{57b} \\
   \textrm{Tr}\left\{\widehat{\widetilde{\mathcal{U}}}(\vec{\lambda},\vec{\mu})\widehat{\widetilde{\mathcal{U}}}^{+ \hspace{-0.45em}+}(\vec{\lambda}',\vec{\mu}')\right\} &= \left(\frac{2\pi}{\hbar}\right)^{3}\delta(\vec{\lambda}-\vec{\lambda}')\delta(\vec{\mu}-\vec{\mu}')  \label{57c}
\end{align}
\end{subequations}
(compare with Eqs. (2.10) and (2.14) of \cite{Przanowski2019}).

First equalities of (\ref{56a}) and (\ref{57a}) say simply that $\widehat{\widetilde{\mathcal{D}}}(k,l)$ and $\widehat{\widetilde{\mathcal{U}}}(\vec{\lambda},\vec{\mu})$ are generalized unitary operators. With the use of these two operators one defines the \textit{generalized Stratonovich-Weyl quantizer} (the \textit{generalized Fano operators}) in close analogy to the ``usual'' Stratonovich-Weyl quantizer (the Fano operators) introduced in  previous works \cite{Przanowski2019, Przanowski2017}. One puts
$$
\widehat{\widetilde{\Omega}}[\mathcal{P},\mathcal{K}](\vec{p},\vec{x},\phi_{m},n):=\left(\frac{\hbar}{2\pi}\right)^{3}\frac{1}{3}\sum_{k,l=0}^{2}\int_{{\mathbb R}^6} d^{3}\lambda d^{3}\mu \,
\mathcal{P}\left(\frac{\hbar \vec{\lambda}\cdot\vec{\mu}}{2}\right)\mathcal{K}\left(\frac{\pi k l}{3}\right)
$$
$$
\times
\exp\left\{-i(\vec{\lambda}\cdot \vec{p} +\vec{\mu}\cdot \vec{x})\right\}\exp\left\{-i(k \cdot \phi_{m}+\phi_{l}\cdot n)\right\}
\widehat{\widetilde{\mathcal{U}}}(\vec{\lambda},\vec{\mu})\widehat{\widetilde{\mathcal{D}}}(k,l)
$$
\begin{equation}\label{58}
=\widehat{H}^{1/2}\,\widehat{\Omega}[\mathcal{P},\mathcal{K}](\vec{p},\vec{x},\phi_{m},n)\,\widehat{H}^{-1/2},
\end{equation}
where $\mathcal{P}\left(\frac{\hbar \vec{\lambda}\cdot\vec{\mu}}{2}\right)$ and $\mathcal{K}\left(\frac{\pi k l}{3}\right)$ are the \textit{kernels} which have been introduced and analysed in detail in \cite{Przanowski2019, Przanowski2017} and $\widehat{\Omega}[\mathcal{P},\mathcal{K}]$ is the Stratonovich-Weyl quantizer defined by Eq. (2.33) in \cite{Przanowski2019}.

One easily finds that the generalized Stratonovich-Weyl quantizer obeys the rules
\begin{subequations}\label{59}
\be
   \widehat{\widetilde{\Omega}}^{+ \hspace{-0.45em}+}[\mathcal{P},\mathcal{K}]=\widehat{\widetilde{\Omega}}[\mathcal{P},\mathcal{K}], \label{59a}
   \ee
   \be
   \textrm{Tr}\left\{\widehat{\widetilde{\Omega}}[\mathcal{P},\mathcal{K}]\right\}=1, \label{59b}
   \ee 
   \be
   \textrm{Tr}\left\{\widehat{\widetilde{\Omega}}[\mathcal{P},\mathcal{K}](\vec{p},\vec{x},\phi_{m},n)\widehat{\widetilde{\Omega}}[\mathcal{P},\mathcal{K}](\vec{p}\,',\vec{x}\,',\phi_{m'},n')\right\} = \nonumber 
  \ee 
   \be
     \left(\frac{\hbar}{2\pi}\right)^{3}\frac{1}{3}\sum_{k,l=0}^{2}\int_{{\mathbb R}^6} d^{3}\lambda d^{3}\mu 
   \left|\mathcal{P}\left(\frac{\hbar \vec{\lambda}\cdot\vec{\mu}}{2}\right)\mathcal{K}\left(\frac{\pi k l}{3}\right)\right|^{2}  \nonumber
   \ee
\be
 \times  \exp\left\{i[\vec{\lambda}\cdot (\vec{p}-\vec{p}\,')+\vec{\mu}\cdot(\vec{x}-\vec{x}\,')]\right\}\exp\left\{i[k \cdot (\phi_{m}-\phi_{m'})+\phi_{l}\cdot (n-n')]\right\}.
     \label{59c}
     \ee
\end{subequations}
From (\ref{59c}) we infer that
\begin{equation}\label{510}
  \textrm{Tr}\left\{\widehat{\widetilde{\Omega}}[\mathcal{P},\mathcal{K}](\vec{p},\vec{x},\phi_{m},n) \widehat{\widetilde{\Omega}}[\mathcal{P},\mathcal{K}](\vec{p}\,',\vec{x}\,',\phi_{m'},n')\right\}=(2\pi\hbar)^{3}3 \,\delta(\vec{p}-\vec{p}')\delta(\vec{x}-\vec{x}')\delta_{mm'}\delta_{nn'}
\end{equation}
for
\begin{equation}\label{511}
  \left|\mathcal{P}\right|=1 \quad \textrm{and} \quad \left|\mathcal{K}\right|=1.
\end{equation}
Of course  the relation (\ref{59a}) says that  the operator $\widehat{\widetilde{\Omega}}[\mathcal{P},\mathcal{K}](\vec{p},\vec{x},\phi_{m},n)$ is a generalized Hermitian operator for every $(\vec{p},\vec{x},\phi_{m},n)\in \Gamma$.

It is convenient to extend  the transformation (\ref{417}) on an arbitrary linear operator. So given an operator $\widehat{\gamma}$ we define $\widehat{\gamma}_{H}$ as
\begin{equation}\label{512}
  \widehat{\gamma}_{H}:=\widehat{H}^{-1/2}\,\widehat{{\gamma}}\,\widehat{H}^{1/2}.
\end{equation}
Comparing (\ref{511}) with (\ref{55}) and (\ref{58}) one gets
\begin{equation}\label{513}
  \widehat{\widetilde{\mathcal{D}}}_{H}=\widehat{\mathcal{D}}, \quad \widehat{\widetilde{\mathcal{U}}}_{H}=\widehat{\mathcal{U}}, \quad \widehat{\widetilde{\Omega}}_{H}[\mathcal{P},\mathcal{K}]=\widehat{\Omega}[\mathcal{P},\mathcal{K}].
\end{equation}
Now we are in position to define the generalized Weyl correspondence between the functions on phase space $\Gamma$ and operators in the Hilbert space $\mathcal{H}$.

According to this correspondence for a function $f=f(\vec{p},\vec{x},\phi_{m},n)$ on $\Gamma$ we assign an operator $\widehat{f}$ in $\mathcal{H}$ given by
\begin{equation}\label{514}
  \widehat{f}=\frac{1}{3(2\pi\hbar)^{3}}\sum_{m,n=0}^{2}\int_{{\mathbb R}^6} d^{3}p d^{3}x f(\vec{p},\vec{x},\phi_{m},n)\widehat{\widetilde{\Omega}}[\mathcal{P},\mathcal{K}](\vec{p},\vec{x},\phi_{m},n)
\end{equation}
(compare with (2.32b) of Ref. \cite{Przanowski2019}). Multiplying both sides of (\ref{514}) by $\widehat{\widetilde{\Omega}}[\mathcal{P},\mathcal{K}](\vec{p}\,',\vec{x}\,',\phi_{m'},n')$, taking the trace and using also (\ref{59c}) one gets the formula inverse to (\ref{514}) as
$$
f(\vec{p},\vec{x},\phi_{m},n)=\frac{1}{9(2\pi)^{6}}\sum_{k,l,m',n'=0}^{2}\int_{{\mathbb R}^{12}} d^{3}\lambda d^{3}\mu d^{3}p' d^{3}x'
$$
$$
\times \left|\mathcal{P}\left(\frac{\hbar \vec{\lambda}\cdot\vec{\mu}}{2}\right)\mathcal{K}\left(\frac{\pi k l}{3}\right)\right|^{-2}
\exp\left\{i[\vec{\lambda}\cdot (\vec{p}-\vec{p}\,')+\vec{\mu}\cdot(\vec{x}-\vec{x}\,')]\right\}
$$
\begin{equation}\label{515}
  \times \exp\left\{i[k \cdot (\phi_{m}-\phi_{m'})+\phi_{l}\cdot(n-n')]\right\}
 \textrm{Tr}\left\{\widehat{f}\,\widehat{\widetilde{\Omega}}[\mathcal{P},\mathcal{K}](\vec{p}\,',\vec{x}\,',\phi_{m'},n')\right\}
\end{equation}
(compare with (2.39) in \cite{Przanowski2019}). 

One easily finds that if (\ref{511}) holds true then Eq. (\ref{515}) simplifies considerably and it reads then
\begin{equation}\label{516}
  f(\vec{p},\vec{x},\phi_{m},n)=\textrm{Tr}\left\{\widehat{f}\,\widehat{\widetilde{\Omega}}[\mathcal{P},\mathcal{K}](\vec{p},\vec{x},\phi_{m},n)\right\}.
\end{equation}
Thus  formulae (\ref{514}) and (\ref{515}) give a one -- to -- one correspondence between  functions on phase space $\Gamma$ and  operators in the Hilbert space $\mathcal{H}$. Employing (\ref{59a}) we conclude that  \textit{operator $\widehat{f}$ is a generalized Hermitian operator, $\widehat{f}^{+ \hspace{-0.45em}+}=\widehat{f}$, iff $f=f(\vec{p},\vec{x},\phi_{m},n)$ is a real function}.

Observe that Eqs. (\ref{514}), (\ref{515}) and (\ref{516}) can be equivalently rewritten by the substitutions:
\begin{equation}
  \widehat{f} \longmapsto \widehat{f}_{H}=\widehat{H}^{-1/2}\widehat{f}\widehat{H}^{1/2} \quad \textrm{and} \quad \widehat{\widetilde{\Omega}}[\mathcal{P},\mathcal{K}] \longmapsto \widehat{\Omega}[\mathcal{P},\mathcal{K}], \nonumber
\end{equation}
where $\widehat{\Omega}[\mathcal{P},\mathcal{K}]$ is the Stratonovich-Weyl quantizer introduced in Ref. \cite{Przanowski2019}.

The notion of star product can be introduced in the same way as it is done in nonrelativistic case.
Namely, if $f=f(\vec{p},\vec{x},\phi_{m},n)$ and $g=g(\vec{p},\vec{x},\phi_{m},n)$ are functions on $\Gamma$, and $\widehat{f}$ and $\widehat{g}$ are the respective operators in $\mathcal{H}$ then the function corresponding to the product $\widehat{f}\cdot\widehat{g}$ is denoted by $f \ast g$ and according to (\ref{515}) it reads
$$
(f \ast g)(\vec{p},\vec{x},\phi_{m},n)=\frac{1}{9(2\pi)^{6}}\sum_{k,l,m',n'=0}^{2}\int_{{\mathbb R}^6} d^{3}\lambda d^{3}\mu d^{3}p' d^{3}x'
$$
$$
\left|\mathcal{P}\left(\frac{\hbar \vec{\lambda}\cdot\vec{\mu}}{2}\right)\mathcal{K}\left(\frac{\pi k l}{3}\right)\right|^{-2}
\exp\left\{i[\vec{\lambda}\cdot (\vec{p}-\vec{p}\,')+\vec{\mu}\cdot(\vec{x}-\vec{x}\,')]\right\}
$$
\begin{equation}\label{517}
\times
 \exp\left\{i[k \cdot (\phi_{m}-\phi_{m'})+\phi_{l}\cdot(n-n')]\right\}
 \textrm{Tr}\left\{\widehat{f}\cdot\widehat{g}\,\widehat{\widetilde{\Omega}}[\mathcal{P},\mathcal{K}](\vec{p}\,',\vec{x}\,',\phi_{m'},n')\right\}.
\end{equation}
Inserting into (\ref{517}) $\widehat{f}$ and $\widehat{g}$ in accordance with (\ref{514}) one gets
$$
(f \ast g)(\vec{p},\vec{x},\phi_{m},n)=\frac{1}{81 \hbar^{6}(2\pi)^{12}}\sum_{k,l,m',n', m'',n'',m''',n'''=0}^{2}\int_{{\mathbb R}^{24}} d^{3}\lambda d^{3}\mu d^{3}p' d^{3}x'd^{3}p'' d^{3}x''d^{3}p''' d^{3}x'''
$$
$$
\times \left|\mathcal{P}\left(\frac{\hbar \vec{\lambda}\cdot\vec{\mu}}{2}\right)\mathcal{K}\left(\frac{\pi k l}{3}\right)\right|^{-2}
\exp\left\{i[\vec{\lambda}\cdot (\vec{p}-\vec{p}\,')+\vec{\mu}\cdot(\vec{x}-\vec{x}\,')]\right\}
$$
$$
\times
\exp\left\{i[k \cdot (\phi_{m}-\phi_{m'})+\phi_{l}\cdot(n-n')]\right\} f(\vec{p}\,'',\vec{x}\,'',\phi_{m''},n'')
$$
\begin{equation}\label{518}
\times
 \textrm{Tr}\left\{\widehat{\widetilde{\Omega}}[\mathcal{P},\mathcal{K}](\vec{p}\,',\vec{x}\,',\phi_{m'},n')
 \widehat{\widetilde{\Omega}}[\mathcal{P},\mathcal{K}](\vec{p}\,'',\vec{x}\,'',\phi_{m''},n'')
 \widehat{\widetilde{\Omega}}[\mathcal{P},\mathcal{K}](\vec{p}\,''',\vec{x}\,''',\phi_{m'''},n''')\right\}g(\vec{p}\,''',\vec{x}\,''',\phi_{m'''},n''').
\end{equation}
Using the relation between $\widehat{\widetilde{\Omega}}[\mathcal{P},\mathcal{K}]$ and $\widehat{\Omega}[\mathcal{P},\mathcal{K}]$ given by (\ref{58}) we quickly conclude that in (\ref{518}) one can equivalently put $\widehat{\Omega}[\mathcal{P},\mathcal{K}]$ instead of $\widehat{\widetilde{\Omega}}[\mathcal{P},\mathcal{K}]$.
Therefore the star product given by (\ref{518}) is exactly the same as the star product defined in our previous work \cite{Przanowski2019} (see Eq. (3.2) in \cite{Przanowski2019}) for $(s+1)=3$. An explicit expression of product \eqref{518} is presented in Appendix \ref{appendixA}.

If the kernels $\mathcal{P}$ and $\mathcal{K}$ fulfill the conditions (\ref{511}) then $f \ast g$ can be written in a simpler form
$$
(f \ast g)(\vec{p},\vec{x},\phi_{m},n)=\frac{1}{9(2\pi\hbar)^{6}}\sum_{m',n', m'',n''=0}^{2}\int_{{\mathbb R}^{12}} d^{3}p' d^{3}x'd^{3}p'' d^{3}x'' f(\vec{p}\,',\vec{x}\,',\phi_{m'},n')
$$
\begin{equation}\label{519}
\times
 \textrm{Tr}\left\{\widehat{\widetilde{\Omega}}[\mathcal{P},\mathcal{K}](\vec{p},\vec{x},\phi_{m},n)
 \widehat{\widetilde{\Omega}}[\mathcal{P},\mathcal{K}](\vec{p}\,',\vec{x}\,',\phi_{m'},n')
 \widehat{\widetilde{\Omega}}[\mathcal{P},\mathcal{K}](\vec{p}\,'',\vec{x}\,'',\phi_{m''},n'')\right\}g(\vec{p}\,'',\vec{x}\,'',\phi_{m''},n''),
\end{equation}
where due to the above comment  we put $\widehat{\Omega}[\mathcal{P},\mathcal{K}]$ in place of $\widehat{\widetilde{\Omega}}[\mathcal{P},\mathcal{K}]$.

We then define a photon Wigner function. Assume that operator $\widehat{f}$ satisfying $\widehat{f}^{+ \hspace{-0.45em}+}=\widehat{f}$, represents a photon observable. The average value of this observable in the photon state $\widehat{\rho}$ is determined by the formula (\ref{426}) with $\widehat{\vartheta}$ substituted by $\widehat{f}$.

Using also (\ref{514}) one finally gets
\begin{align}\label{520}
 \langle \widehat{f}\rangle&=\textrm{Tr}\left\{\widehat{f}\widehat{\rho}\right\}=\textrm{Tr}\left\{\widehat{f}_{H}\widehat{\rho}_{H}\right\} \nonumber \\
&=\frac{1}{3(2\pi\hbar)^{3}}\sum_{m,n=0}^{2}\int_{{\mathbb R}^6} d^{3}p d^{3}x f(\vec{p},\vec{x},\phi_{m},n)
\textrm{Tr}\left\{\widehat{\rho}_{H}\widehat{\Omega}[\mathcal{P},\mathcal{K}](\vec{p},\vec{x},\phi_{m},n)\right\} \nonumber \\
&=\frac{1}{3(2\pi\hbar)^{3}}\sum_{m,n=0}^{2}\int_{{\mathbb R}^6} d^{3}p d^{3}x f(\vec{p},\vec{x},\phi_{m},n)
\textrm{Tr}\left\{\widehat{\rho}\,\widehat{\widetilde{\Omega}}[\mathcal{P},\mathcal{K}](\vec{p},\vec{x},\phi_{m},n)\right\}.
\end{align}
In close analogy to the previous works we define the \textit{Wigner function of photon in  state $\widehat{\rho}$ for the kernels $(\mathcal{P},\mathcal{K})$} as
\begin{align}\label{521}
 \rho_{W}[\mathcal{P},\mathcal{K}](\vec{p},\vec{x},\phi_{m},n):&=\frac{1}{3(2\pi\hbar)^{3}}
 \textrm{Tr}\left\{\widehat{\rho}\,\widehat{\widetilde{\Omega}}[\mathcal{P},\mathcal{K}](\vec{p},\vec{x},\phi_{m},n)\right\} \nonumber \\
 &=\frac{1}{3(2\pi\hbar)^{3}}\textrm{Tr}\left\{\widehat{\rho}_{H}\, \widehat{\Omega}[\mathcal{P},\mathcal{K}](\vec{p},\vec{x},\phi_{m},n)\right\}.
\end{align}
Formula inverse to (\ref{521}) reads
$$
\widehat{\rho}=\frac{1}{9 (2\pi)^{6}}\sum_{k,l,m,n,m',n'=0}^{2}\int_{{\mathbb R}^{18}} d^{3}\lambda d^{3}\mu d^{3}p d^{3}x d^{3}p' d^{3}x'
\left|\mathcal{P}\left(\frac{\hbar \vec{\lambda}\cdot\vec{\mu}}{2}\right)\mathcal{K}\left(\frac{\pi k l}{3}\right)\right|^{-2}
$$
$$
\times
\exp\left\{i[\vec{\lambda}\cdot (\vec{p}-\vec{p}\,')+\vec{\mu}\cdot(\vec{x}-\vec{x}\,')]\right\} \exp\left\{i[k \cdot (\phi_{m}-\phi_{m'})+\phi_{l}\cdot(n-n')]\right\}
$$
\begin{equation}\label{522}
\times
 \rho_{W}[\mathcal{P},\mathcal{K}](\vec{p},\vec{x},\phi_{m},n)\widehat{\widetilde{\Omega}}[\mathcal{P},\mathcal{K}](\vec{p}\,',\vec{x}\,',\phi_{m'},n').
\end{equation}
Eq. (\ref{522}) simplifies considerably when the conditions (\ref{511}) are fulfilled. In that case we have
\begin{equation}\label{523}
 \widehat{\rho}= \sum_{m,n=0}^{2}\int_{{\mathbb R}^{6}} d^{3}p d^{3}x \rho_{W}[\mathcal{P},\mathcal{K}](\vec{p},\vec{x},\phi_{m},n)\widehat{\widetilde{\Omega}}[\mathcal{P},\mathcal{K}](\vec{p},\vec{x},\phi_{m},n).
\end{equation}
One quickly finds that the Wigner function (\ref{521}) has the following properties
\begin{subequations}\label{524}
\begin{align}
   \rho_{W}^{\ast}[\mathcal{P},\mathcal{K}]=\rho_{W}[\mathcal{P},\mathcal{K}], \;\;\;\;\;\;\;\;\;\;\;\;\;\;\;\;\;\;\;\;\;\;\;\;\;\;\;\;\;\;\;\;\;\;\; \label{524a}\\
   \sum_{m,n=0}^{2}\int_{{\mathbb R}^{6}} d^{3}p d^{3}x \rho_{W}[\mathcal{P},\mathcal{K}](\vec{p},\vec{x},\phi_{m},n)=\textrm{Tr}\left\{\widehat{\rho}\right\}=1, \;\;\;\;\;\;\;\;\;\;\;\;\; \label{524b} \\
   \sum_{m,n=0}^{2}\int_{{\mathbb R}^{3}} d^{3}p \rho_{W}[\mathcal{P},\mathcal{K}](\vec{p},\vec{x},\phi_{m},n)=\textrm{Tr}\left\{\widehat{\rho}_{H}|\vec{x}\rangle\langle \vec{x}|\right\}=
   \textrm{Tr}\left\{\widehat{\rho}\widehat{H}^{1/2}|\vec{x}\rangle\langle \vec{x}|\widehat{H}^{-1/2}\right\},  \label{524c}
\end{align}
\end{subequations}
$$
   \sum_{m,n=0}^{2}\int_{{\mathbb R}^{3}} d^{3}x \rho_{W}[\mathcal{P},\mathcal{K}](\vec{p},\vec{x},\phi_{m},n) = \textrm{Tr}\left\{\widehat{\rho}_{H}|\vec{p}\,\rangle\langle \vec{p}\,|\right\}=
   \textrm{Tr}\left\{\widehat{\rho}|\vec{p}\,\rangle\langle \vec{p}\,|\right\},
$$
$$
   \sum_{m=0}^{2}\int_{{\mathbb R}^{6}} d^{3}p d^{3}x \rho_{W}[\mathcal{P},\mathcal{K}](\vec{p},\vec{x},\phi_{m},n) = \textrm{Tr}\left\{\widehat{\rho}_{H}|n\rangle\langle n|\right\}=
   \textrm{Tr}\left\{\widehat{\rho}\widehat{H}^{1/2}|n\rangle\langle n|\widehat{H}^{-1/2}\right\},
$$
$$
   \sum_{n=0}^{2}\int_{{\mathbb R}^{6}} d^{3}p d^{3}x \rho_{W}[\mathcal{P},\mathcal{K}](\vec{p},\vec{x},\phi_{m},n) = \textrm{Tr}\left\{\widehat{\rho}_{H}|\phi_{m}\rangle\langle \phi_{m}|\right\}=
   \textrm{Tr}\left\{\widehat{\rho}\widehat{H}^{1/2}|\phi_{m}\rangle\langle \phi_{m}|\widehat{H}^{-1/2}\right\}.
$$
Properties (\ref{524a}) and (\ref{524b}) are fulfilled also by any reasonable Wigner function arising from the Weyl -- Wigner -- Moyal formalism. Properties (\ref{524c}) imitate the marginal probability distributions. However,  interpretation of (\ref{524c}) is more subtle.

Quantity appearing in (\ref{524c}) cannot be interpreted as the probability distribution of  position $\vec{x}$ of photon since  operator $\widehat{\vec{x}}$ is not a photon observable at all. To find a correct interpretation of that expression we first rewrite Eq. (\ref{322}) in  form
\begin{align}\label{525}
 \wp (\vec{x}\in V)&= \frac{\int_{V}d^{3}x |\langle\vec{x}|{\bf \Psi}\rangle|^{2}}{\langle{\bf \Psi}|{\bf \Psi}\rangle} 
 = \frac{\int_{V}d^{3}x \textrm{Tr}\left\{\widehat{H}^{1/2}\,\widehat{\rho}_{H}\,\widehat{H}^{1/2}|\vec{x}\rangle\langle \vec{x}|\right\}}{\textrm{Tr}\left\{\widehat{\rho}_{H}\,\widehat{H}\right\}} \nonumber \\
 &= \frac{\int_{V}d^{3}x \textrm{Tr}\left\{\widehat{\rho}\,\widehat{H}|\vec{x}\rangle\langle \vec{x}|\right\}}{\textrm{Tr}\left\{\widehat{\rho}\widehat{H}\right\}}
\end{align}
where $\widehat{\rho}_{H}$ is defined by (\ref{424}). We assume that the formula (\ref{525}) is valid for any state (pure or mixed) $\widehat{\rho}$.

Substituting in (\ref{525})
\begin{equation}\label{526}
  \widehat{\rho}_{1}:=\frac{\widehat{H}^{-1/2}\widehat{\rho}\widehat{H}^{-1/2}}
  {\textrm{Tr}\left\{\widehat{H}^{-1/2}\widehat{\rho}\widehat{H}^{-1/2} \right\}} \quad \Longrightarrow \quad
  \widehat{\rho}_{1H}=\frac{\widehat{H}^{-1/2}\widehat{\rho}_{H}\widehat{H}^{-1/2}}
  {\textrm{Tr}\left\{\widehat{H}^{-1/2}\widehat{\rho}_{H}\widehat{H}^{-1/2} \right\}}
\end{equation}
in the place of $\widehat{\rho}$ and $\widehat{\rho}_{H}$, respectively, one gets
\begin{align}\label{527}
 \wp(\vec{x}\in V)&= \frac{\int_{V}d^{3}x \textrm{Tr}\left\{\widehat{\rho}_{1}\widehat{H}|\vec{x}\rangle\langle \vec{x}|\right\}}{\textrm{Tr}\left\{\widehat{\rho}_{1}\cdot\widehat{H}\right\}} 
= \int_{V}d^{3}x \textrm{Tr}\left\{\widehat{\rho}_{H}|\vec{x}\rangle\langle \vec{x}|\right\}.
\end{align}
This last formula provides us with simple interpretation for the first of Eqs. (\ref{524c}). Namely, \textit{$\textrm{Tr}\left\{\widehat{\rho}_{H}|\vec{x}\rangle\langle \vec{x}|\right\}$ is the density of probability to find the photon energy localized at  point $\vec{x}$ for  state $\widehat{\rho}_{1}$ given by (\ref{526})}.

Interpretation of the second formula of Eqs. (\ref{524c}) is much simpler and clear. Indeed, the \textit{function $\textrm{Tr}\left\{\widehat{\rho}|\vec{p}\,\rangle\langle \vec{p}\,|\right\}$ defines the probability distribution for the momentum of photon $\vec{p}=\hbar\vec{k}$ in  state $\widehat{\rho}$}.

Sense of  last two properties of (\ref{524c}) is rather unclear. Although  operators $\widehat{H}^{1/2}|n\rangle\langle n|\widehat{H}^{-1/2}$, $n=0,1,2$, and $\widehat{H}^{1/2}|\phi_{m}\rangle\langle \phi_{m}|\widehat{H}^{-1/2}$, $m=0,1,2$, are generalized Hermitian operators, they are not  photon observables in general. So we are not able to give any reasonable  explanation of these two formulae.

Now we are going to find an evolution equation for the Wigner function $\rho_{W}[\mathcal{P},\mathcal{K}](\vec{p},\vec{x},\phi_{m},n;t)$. From (\ref{521}), using also the Liouville -- von Neumann equation (\ref{427}) one gets
\begin{equation}\label{528}
 \frac{\partial \rho_{W}[\mathcal{P},\mathcal{K}]}{\partial t}+\frac{1}{3(2\pi\hbar)^{3}}\textrm{Tr}\left\{\frac{1}{i \hbar} \left[\widehat{\rho},\widehat{H}\right]\widehat{\widetilde{\Omega}}[\mathcal{P},\mathcal{K}]\right\}=0
\end{equation}
Employing (\ref{514}), (\ref{522}), (\ref{518}) and defining
$$
R_{W}[\mathcal{P},\mathcal{K}](\vec{p}\,',\vec{x}\,',\phi_{m'},n';t):=\frac{1}{9 (2\pi)^{6}}\sum_{k,l,m,n=0}^{2}\int_{{\mathbb R}^{12}} d^{3}\lambda d^{3}\mu d^{3}p d^{3}x
\left|\mathcal{P}\left(\frac{\hbar \vec{\lambda}\cdot\vec{\mu}}{2}\right)\mathcal{K}\left(\frac{\pi k l}{3}\right)\right|^{-2}
$$
$$
\exp\left\{i[\vec{\lambda}\cdot (\vec{p}-\vec{p}\,')+\vec{\mu}\cdot(\vec{x}-\vec{x}\,')]\right\} \exp\left\{i[k \cdot (\phi_{m}-\phi_{m'})+\phi_{l}\cdot(n-n')]\right\}
$$
\begin{equation}\label{529}
\times \rho_{W}[\mathcal{P},\mathcal{K}](\vec{p},\vec{x},\phi_{m},n;t)
\end{equation}
we can rewrite Eq. (\ref{528}) as
\begin{equation}\label{530}
 \frac{\partial \rho_{W}[\mathcal{P},\mathcal{K}]}{\partial t}+\frac{1}{i \hbar}\Big( R_{W}[\mathcal{P},\mathcal{K}]\ast H - H \ast R_{W}[\mathcal{P},\mathcal{K}] \Big)=0.
\end{equation}
Eq. (\ref{530}) will be called the \textit{Liouville -- von Neumann-Wigner equation}. It is rather involved, but when the kernels $(\mathcal{P},\mathcal{K})$ satisfy  conditions (\ref{511}) then $R_{W}[\mathcal{P},\mathcal{K}]=\rho_{W}[\mathcal{P},\mathcal{K}]$ and Eq. (\ref{530}) takes the simple form
\begin{align}\label{531}
 \frac{\partial \rho_{W}[\mathcal{P},\mathcal{K}]}{\partial t}+\frac{1}{i \hbar}\left( \rho_{W}[\mathcal{P},\mathcal{K}]\ast H - H \ast \rho_{W}[\mathcal{P},\mathcal{K}] \right)=0 \\
 |\mathcal{P}|=1\quad , \quad |\mathcal{K}|=1. \;\;\;\;\;\;\;\;\;\;\;\;\;\;\;\;\;\;\;\;\;\;\;\;\;\; \nonumber
\end{align}
 Hamiltonian $H$ in (\ref{530}) and (\ref{531}) reads
\begin{equation}\label{532}
  H=c\sqrt{\vec{p}\cdot\vec{p}}=c\sqrt{p_{1}^{2}+p_{2}^{2}+p_{3}^{2}}.
\end{equation}
As has been shown in the previous paper \cite{Przanowski2019} one can modify the Liouville -- von Neumann -- Wigner equation so that its form is independent of the kernels $(\mathcal{P},\mathcal{K})$. In the present case, \textit{mutatis mutandi}, we can also develop such an approach.

To this end we define the following correspondence between operators and functions
$$
\widetilde{\widetilde{f}}=\widetilde{\widetilde{f}}(\vec{\lambda},\vec{\mu},k,l) \longleftrightarrow \widehat{f}, \quad \vec{\lambda},\vec{\mu} \in \mathbb{R}^{3}, \enspace k,l=0,1,2,
$$
$$
\widetilde{\widetilde{f}}(\vec{\lambda},\vec{\mu},k,l) = \left(\frac{\hbar}{2\pi}\right)^{3}\frac{1}{3} \textrm{Tr}\left\{\widehat{f}\,\widehat{\widetilde{\mathcal{U}}}^{+ \hspace{-0.45em}+}(\vec{\lambda},\vec{\mu})\widehat{\widetilde{\mathcal{D}}}^{+ \hspace{-0.45em}+}(k,l)\right\},
$$
\begin{equation}\label{533}
\widehat{f} = \sum_{k,l=0}^{2}\int_{{\mathbb R}^6} d^{3}\lambda d^{3}\mu \widetilde{\widetilde{f}}(\vec{\lambda},\vec{\mu},k,l) \widehat{\widetilde{\mathcal{U}}}(\vec{\lambda},\vec{\mu})\widehat{\widetilde{\mathcal{D}}}(k,l).
\end{equation}
A relation between $\widetilde{\widetilde{f}}(\vec{\lambda},\vec{\mu},k,l)$ and $f(\vec{p},\vec{x},\phi_{m},n)$ given in (\ref{514}) and (\ref{515}) reads
$$
f(\vec{p},\vec{x},\phi_{m},n)=\sum_{k,l=0}^{2}\int_{{\mathbb R}^6} d^{3}\lambda d^{3}\mu \left(\mathcal{P}\left(\frac{\hbar \vec{\lambda}\cdot\vec{\mu}}{2}\right)
\mathcal{K}\left(\frac{\pi k l}{3}\right)\right)^{-1} 
$$
\begin{equation}\label{534}
\times
\exp\left\{i(\vec{\lambda}\cdot \vec{p}+\vec{\mu}\cdot \vec{x})\right\} \exp\left\{i(k \phi_{m}+\phi_{l}n)\right\}\widetilde{\widetilde{f}}(\vec{\lambda},\vec{\mu},k,l).
\end{equation}
Then one finds that if $\widetilde{\widetilde{f}} \longleftrightarrow \widehat{f}$ and $\widetilde{\widetilde{g}} \longleftrightarrow \widehat{g}$ then $\widetilde{\widetilde{f}}\boxtimes\widetilde{\widetilde{g}} \longleftrightarrow \widehat{f}\cdot\widehat{g},$ where
$$
(\widetilde{\widetilde{f}}\boxtimes\widetilde{\widetilde{g}})(\vec{\lambda},\vec{\mu},k,l):= \left(\frac{\hbar}{2\pi}\right)^{3}\frac{1}{3} \textrm{Tr}\left\{\widehat{f}\cdot\widehat{g}\;\;\widehat{\widetilde{\mathcal{U}}}^{+ \hspace{-0.45em}+}(\vec{\lambda},\vec{\mu})\widehat{\widetilde{\mathcal{D}}}^{+ \hspace{-0.45em}+}(k,l)\right\}
$$
$$
= \left(\frac{\hbar}{2\pi}\right)^{3}\frac{1}{3}\sum_{k',l',k'',l''=0}^2 \int_{{\mathbb R}^{12}} d^{3}\lambda'd^{3}\mu'd^{3}\lambda''d^{3}\mu'' \widetilde{\widetilde{f}}(\vec{\lambda}',\vec{\mu}',k',l')
$$
$$
\times \textrm{Tr}\left\{\widehat{\widetilde{\mathcal{U}}}^{+ \hspace{-0.45em}+}(\vec{\lambda},\vec{\mu})\widehat{\widetilde{\mathcal{D}}}^{+ \hspace{-0.45em}+}(k,l)\widehat{\widetilde{\mathcal{U}}}
 (\vec{\lambda}',\vec{\mu}')\widehat{\widetilde{\mathcal{D}}}(k',l')\widehat{\widetilde{\mathcal{U}}}(\vec{\lambda}'',\vec{\mu}'')
 \widehat{\widetilde{\mathcal{D}}}(k'',l'')\right\}\widetilde{\widetilde{g}}(\vec{\lambda}'',\vec{\mu}'',k'',l'')
$$
$$
= \left(\frac{\hbar}{2\pi}\right)^{3}\frac{1}{3}\sum_{k',l',k'',l''=0}^2 \int_{{\mathbb R}^{12}} d^{3}\lambda'd^{3}\mu'd^{3}\lambda''d^{3}\mu'' \widetilde{\widetilde{f}}(\vec{\lambda}',\vec{\mu}',k',l')
$$
\begin{equation}\label{535}
\times \textrm{Tr}\left\{\widehat{\mathcal{U}}^{\dag}(\vec{\lambda},\vec{\mu})\widehat{\mathcal{D}}^{\dag}(k,l)\widehat{\mathcal{U}}
 (\vec{\lambda}',\vec{\mu}')\widehat{\mathcal{D}}(k',l')\widehat{\mathcal{U}}(\vec{\lambda}'',\vec{\mu}'')
 \widehat{\mathcal{D}}(k'',l'')\right\}\widetilde{\widetilde{g}}(\vec{\lambda}'',\vec{\mu}'',k'',l'').
\end{equation}
Similarly as $\widetilde{\widetilde{f}}$ in (\ref{533}) we define $\widetilde{\widetilde{\rho}}_W$
\begin{equation}\label{536}
  \widetilde{\widetilde{\rho}}_{W}(\vec{\lambda},\vec{\mu},k,l;t):=\left(\frac{\hbar}{2\pi}\right)^{3}\frac{1}{3} \textrm{Tr}\left\{\widehat{\rho}(t)\widehat{\widetilde{\mathcal{U}}}^{+ \hspace{-0.45em}+}(\vec{\lambda},\vec{\mu})\widehat{\widetilde{\mathcal{D}}}(k,l)\right\}
\end{equation}
$$
\Longrightarrow \quad \widehat{\rho}(t)= \sum_{k,l=0}^{2}\int d^{3}\lambda d^{3}\mu \widetilde{\widetilde{\rho}}_{W}(\vec{\lambda},\vec{\mu},k,l;t) \widehat{\widetilde{\mathcal{U}}}(\vec{\lambda},\vec{\mu})\widehat{\widetilde{\mathcal{D}}}(k,l).
$$
Differentiating $\widetilde{\widetilde{\rho}}_{W}(\vec{\lambda},\vec{\mu},k,l;t)$ over $t$, using then the Liouville -- von Neumann equation (\ref{427}) and the definition of the $\boxtimes$-product (\ref{535}) one finds the evolution equation for $\widetilde{\widetilde{\rho}}_{W}$
\begin{equation}\label{537}
 \frac{\partial \widetilde{\widetilde{\rho}}_{W}}{\partial t}+\frac{1}{i \hbar}\left( \widetilde{\widetilde{\rho}}_{W}\boxtimes \widetilde{\widetilde{H}} - \widetilde{\widetilde{H}} \boxtimes \widetilde{\widetilde{\rho}}_{W} \right)=0.
\end{equation}
This equation we also call the \textit{Liouville -- von Neumann-Wigner equation}. Given any solution of Eq. (\ref{537}) one finds the respective Wigner function $\rho_{W}[\mathcal{P},\mathcal{K}](\vec{p},\vec{x},\phi_{m},n;t)$ from  the formula
$$
\rho_{W}[\mathcal{P},\mathcal{K}](\vec{p},\vec{x},\phi_{m},n;t)=\frac{1}{(2\pi\hbar)^{3}}\frac{1}{3}\sum_{k,l=0}^{2}\int_{{\mathbb R}^6} d^{3}\lambda d^{3}\mu \mathcal{P}^{\ast}\left(\frac{\hbar \vec{\lambda}\cdot\vec{\mu}}{2}\right)\mathcal{K}^{\ast}\left(\frac{\pi k l}{3}\right)
$$
\begin{equation}\label{538}
\times
\exp\left\{i(\vec{\lambda}\cdot \vec{p}+\vec{\mu}\cdot \vec{x})\right\} \exp\left\{i(k \phi_{m}+\phi_{l}n)\right\}\widetilde{\widetilde{\rho}}_{W}(\vec{\lambda},\vec{\mu},k,l;t).
\end{equation}
\section{Some explicit forms of the photon Wigner functions for specific kernels}
\label{sec6}
\setcounter{equation}{0}

This section contains expressions for quantities appearing in the phase space description of photons. They result from considerations presented in the previous section. 

Inserting (\ref{58}) with (\ref{53}) and (\ref{54}) into (\ref{521}) one can express the Wigner function in the form
$$
\rho_{W}[\mathcal{P},\mathcal{K}](\vec{p},\vec{x},\phi_{m},n)=\frac{1}{9(2\pi)^{6}}\sum_{k,l,n'=0}^{2}\int_{{\mathbb R}^9} d^{3}\lambda d^{3}\mu d^{3}x' \mathcal{P}\left(\frac{\hbar \vec{\lambda}\cdot\vec{\mu}}{2}\right)\mathcal{K}\left(\frac{\pi k l}{3}\right)
$$
$$
\times
\exp\left\{i \frac{\pi k l}{3}\right\}\exp\left\{-i[\vec{\lambda}\cdot \vec{p}+\vec{\mu}\cdot(\vec{x}-\vec{x}\,')]\right\}\exp\left\{-i[k \phi_{m}+\phi_{l}(n-n')]\right\}
$$
$$
\times 
 \Big< \vec{x}\,'+ \frac{\hbar\vec{\lambda}}{2}, (n'+k)\; mod \, 3 \Big|\widehat{\rho}_{H} \Big|\vec{x}\,'-\frac{\hbar\vec{\lambda}}{2}, n' \Big>
$$
$$
=\frac{1}{9(2\pi)^{6}}\sum_{k,l,n'=0}^{2}\int_{{\mathbb R}^9} d^{3}\lambda d^{3}\mu d^{3}p' \mathcal{P}\left(\frac{\hbar \vec{\lambda}\cdot\vec{\mu}}{2}\right)\mathcal{K}\left(\frac{\pi k l}{3}\right)\exp\left\{i \frac{\pi k l}{3}\right\}
$$
$$
\times \exp\left\{-i[\vec{\lambda}\cdot (\vec{p}-\vec{p}\,')+\vec{\mu}\cdot\vec{x}]\right\}\exp\left\{-i[k \phi_{m}+\phi_{l}(n-n')]\right\}
$$
\begin{equation}\label{61}
  \times \Big< \vec{p}\,'- \frac{\hbar\vec{\mu}}{2}, (n'+k)\; mod \, 3 \Big|\widehat{\rho}_{H} \Big|\vec{p}\,'+\frac{\hbar\vec{\mu}}{2}, n'   \Big>.
\end{equation}
Equivalently, we can write
$$
\rho_{W}[\mathcal{P},\mathcal{K}](\vec{p},\vec{x},\phi_{m},n)=\frac{1}{9(2\pi)^{6}}\sum_{k,l,m'=0}^{2}\int_{{\mathbb R}^9} d^{3}\lambda d^{3}\mu d^{3}x' \mathcal{P}\left(\frac{\hbar \vec{\lambda}\cdot\vec{\mu}}{2}\right)\mathcal{K}\left(\frac{\pi k l}{3}\right)
$$
$$ \times
\exp\left\{i \frac{\pi k l}{3}\right\}\exp\left\{-i[\vec{\lambda}\cdot \vec{p}+\vec{\mu}\cdot(\vec{x}-\vec{x}\,')]\right\}\exp\left\{-i[k (\phi_{m}-\phi_{m'})+\phi_{l}n]\right\}
$$
$$ \times
\Big< \vec{x}\,'+ \frac{\hbar\vec{\lambda}}{2}, \phi_{m'}\Big|\widehat{\rho}_{H}\Big|\vec{x}\,'-\frac{\hbar\vec{\lambda}}{2}, \phi_{m'+l}\Big>
$$
$$
=\frac{1}{9(2\pi)^{6}}\sum_{k,l,m'=0}^{2}\int_{{\mathbb R}^9} d^{3}\lambda d^{3}\mu d^{3}p' \mathcal{P}\left(\frac{\hbar \vec{\lambda}\cdot\vec{\mu}}{2}\right)\mathcal{K}\left(\frac{\pi k l}{3}\right)\exp\left\{i \frac{\pi k l}{3}\right\}
$$
$$ \times
\exp\left\{-i[\vec{\lambda}\cdot (\vec{p}-\vec{p}\,')+\vec{\mu}\cdot\vec{x}\,]\right\}\exp\left\{-i[k \cdot(\phi_{m}-\phi_{m'})+\phi_{l} \cdot n]\right\}
$$
\begin{equation}\label{62} \times
  \Big< \vec{p}\,'- \frac{\hbar\vec{\mu}}{2}, \phi_{m'} \Big|\widehat{\rho}_{H}\Big|\vec{p}\,'+\frac{\hbar\vec{\mu}}{2}, \phi_{m'+l}\Big>.
\end{equation}
Assume now that the photon is in  pure state $|{\bf \Psi}\rangle$, $\langle{\bf \Psi}|{\bf \Psi}\rangle_{BB}=1$. Then $\widehat{\rho}_{H}$ is given by (\ref{424}). Substituting this $\widehat{\rho}_{H}$ into the last formula of (\ref{61}), having also in mind that
\begin{equation}\label{63}
  \langle \vec{p}\,|{\bf \Psi}\rangle=\frac{1}{\hbar^{3/2}}\langle \vec{k}|{\bf \Psi}\rangle=\frac{1}{\hbar}\sqrt{\frac{c}{(2\pi)^{3}}}\widetilde{{\Psi}}(\vec{k}),
\end{equation}
where
\begin{equation}\label{64}
  \widetilde{{\bf \Psi}}(\vec{k})=\left(
                                       \begin{array}{c}
                                         \widetilde{\Psi}_{1}(\vec{k}) \\
                                         \widetilde{\Psi}_{2}(\vec{k})\\
                                         \widetilde{\Psi}_{3}(\vec{k}) \\
                                       \end{array}
                                     \right)
\end{equation}
is defined by (\ref{316}), one gets the Wigner function as
$$
\rho_{W}[\mathcal{P},\mathcal{K}](\vec{p},\vec{x},\phi_{m},n)=\frac{1}{9(2\pi)^{9}}\sum_{k,l,n'=0}^{2}\int_{{\mathbb R}^9} d^{3}\lambda d^{3}\mu d^{3}k' \mathcal{P}\left(\frac{\hbar \vec{\lambda}\cdot\vec{\mu}}{2}\right)\mathcal{K}\left(\frac{\pi k l}{3}\right)\exp\left\{i \frac{\pi k l}{3}\right\}
$$
$$
\times
\exp\left\{-i[\vec{\lambda}\cdot (\vec{p}-\vec{p}\,')+\vec{\mu}\cdot\vec{x}]\right\}\exp\left\{-i[k \phi_{m}+\phi_{l}(n-n')]\right\}
$$
\begin{equation}\label{65} \times
 \left(\left|\vec{k}'-\frac{\vec{\mu}}{2}\right| \left|\vec{k}'+\frac{\vec{\mu}}{2} \right|\right)^{-\frac{1}{2}} \widetilde{\Psi}_{[(n'+k)\; mod \, 3]+1} \left(\vec{k}'-\frac{\vec{\mu}}{2}\right)
 \widetilde{\Psi}_{n'+1}^{\ast} \left(\vec{k}'+\frac{\vec{\mu}}{2} \right).
\end{equation}
Employing then Eq. (\ref{317}) giving the time evolution of photon wave function $\widetilde{{\bf \Psi}}(\vec{k},t)$ we find the time evolution of Wigner function (\ref{65})
$$
\rho_{W}[\mathcal{P},\mathcal{K}](\vec{p},\vec{x},\phi_{m},n;t)=\frac{1}{9(2\pi)^{9}}\sum_{k,l,n'=0}^{2}\int_{{\mathbb R}^9} d^{3}\lambda d^{3}\mu d^{3}k' \mathcal{P}\left(\frac{\hbar \vec{\lambda}\cdot\vec{\mu}}{2}\right)\mathcal{K}\left(\frac{\pi k l}{3}\right)
$$
$$ \times
\frac{\exp\left\{-i\left(|\vec{k}'-\frac{\vec{\mu}}{2}|-|\vec{k}'+\frac{\vec{\mu}}{2}|\right)ct\right\}}
{\left( \left|\vec{k}'-\frac{\vec{\mu}}{2} \right| \left|\vec{k}'+\frac{\vec{\mu}}{2} \right|\right)^{\frac{1}{2}}}\exp\left\{i \frac{\pi k l}{3}\right\}
$$
$$ \times
\exp\left\{-i[\vec{\lambda}\cdot (\vec{p}-\vec{p}\,')+\vec{\mu}\cdot\vec{x}]\right\}\exp\left\{-i[k \phi_{m}+\phi_{l}(n-n')]\right\}
$$
\begin{equation}\label{66} \times
\widetilde{\Psi}_{[(n'+k)\; mod \, 3]+1} \left(\vec{k}'-\frac{\vec{\mu}}{2} \right)\widetilde{\Psi}_{n'+1}^{\ast} \left(\vec{k}'+\frac{\vec{\mu}}{2} \right), \quad \vec{p}'=\hbar\vec{k}' .
\end{equation}

Let us  apply  formula \eqref{66} to the case when
\begin{equation}\label{67}
  \mathcal{P}\left(\frac{\hbar \vec{\lambda}\cdot\vec{\mu}}{2}\right)=1, \quad \mathcal{K}\left(\frac{\pi k l}{3}\right)=(-1)^{kl}.
\end{equation}
As is known from Refs. \cite{Bialynicki1994, Przanowski2017} (see also the references therein) such a choice of kernels is acceptable for $s+1=3$. Inserting (\ref{67}) into (\ref{66}) one gets (we omit the symbol $[\mathcal{P},\mathcal{K}]$ at $\rho_{W}$)
$$
\rho_{W}(\vec{p},\vec{x},\phi_{m},n;t)=\frac{1}{3(2\pi)^{6}\hbar^{3}}\Re \Bigg\{ \int_{{\mathbb R}^3} \frac{d^{3}\mu}{\left( \left|\vec{k}'-\frac{\vec{\mu}}{2} \right| \left|\vec{k}'+\frac{\vec{\mu}}{2} \right|\right)^{\frac{1}{2}}} 
$$
$$ \times
\exp\left\{-i\left( \left|\vec{k}'-\frac{\vec{\mu}}{2}\right|- \left|\vec{k}'+\frac{\vec{\mu}}{2}\right|\right)ct\right\}\exp\left\{-i\vec{\mu}\cdot\vec{x}\right\}
$$
\[
\times 
\left[\widetilde{\Psi}_{n+1}\left(\vec{k}-\frac{\vec{\mu}}{2}\right)\widetilde{\Psi}_{n+1}^{\ast}\left(\vec{k}+\frac{\vec{\mu}}{2}\right) \right.
\]
\begin{equation}\label{68}
\left.
 +2\exp\left\{-i\phi_{m}\right\}
\widetilde{\Psi}_{[(n+2)\; mod \, 3]+1}\left(\vec{k}-\frac{\vec{\mu}}{2}\right)\widetilde{\Psi}_{[(n+1)\; mod \, 3]+1}^{\ast} \left(\vec{k}+\frac{\vec{\mu}}{2}\right)\right]\Bigg\}, \quad \vec{p}=\hbar\vec{k}
\end{equation}
(note that the kernels given by (\ref{67}) fulfill the conditions (\ref{511})).

The next example concerns the choice of kernels $(\mathcal{P},\mathcal{K})$
\begin{equation}\label{69}
  \mathcal{P}\left(\frac{\hbar \vec{\lambda}\cdot\vec{\mu}}{2}\right)=1, \quad \mathcal{K}\left(\frac{\pi k l}{3}\right)=\cos\left(\frac{\pi k l}{3}\right).
\end{equation}
Substituting (\ref{69}) into (\ref{66}) and performing straightforward calculations one gets (we omit the symbol $[\mathcal{P},\mathcal{K}]$ at $\rho_{W}$)
$$
\rho_{W}(\vec{p},\vec{x},\phi_{m},n;t)=\frac{1}{3(2\pi)^{6}\hbar^{3}}\Re\Bigg\{\sum_{n'=0}^{2} \int_{{\mathbb R}^3} \frac{d^{3}\mu}{\left(\left|\vec{k}-\frac{\vec{\mu}}{2}\right| \left|\vec{k}+\frac{\vec{\mu}}{2} \right|\right)^{\frac{1}{2}}}
$$
$$ \times
\exp\left\{-i\left(|\vec{k}-\frac{\vec{\mu}}{2}|-|\vec{k}+\frac{\vec{\mu}}{2}|\right)ct\right\}\exp\left\{-i\vec{\mu}\cdot\vec{x}\right\}
$$
\begin{equation}\label{610} \times 
 \exp\left\{-i\phi_{m}(n-n')\right\}\widetilde{\Psi}_{n+1}\left(\vec{k}-\frac{\vec{\mu}}{2}\right)\widetilde{\Psi}_{n'+1}^{\ast}\left(\vec{k}+\frac{\vec{\mu}}{2} \right)\Bigg\},
\quad \vec{p}=\hbar\vec{k}.
\end{equation}
Constraint equation (\ref{312}) which is equivalent to the condition
\begin{equation}\label{611}
  \sum_{j=1}^{3}k_{j}\widetilde{\Psi}_{j}(\vec{k})=0
\end{equation}
leads to the following constraint equation for $\rho_{W}$ given by (\ref{610})
\begin{equation}\label{612}
  \sum_{n=0}^{2}\left(k_{n+1}-\frac{1}{2}i\frac{\partial}{\partial x^{n+1}}\right) \Delta_{W}(\vec{p},\vec{x},0,n;t)=0,
\end{equation}
$$
\Delta_{W}(\vec{p},\vec{x},\phi_{m},n;t):=\frac{1}{3(2\pi)^{6}\hbar^{3}} \sum_{n'=0}^{2} \int_{{\mathbb R}^3} \frac{d^{3}\mu}{\left(|\vec{k}-\frac{\vec{\mu}}{2}||\vec{k}+\frac{\vec{\mu}}{2}|\right)^{\frac{1}{2}}}
$$
$$ \times
\exp\left\{-i\left(\left|\vec{k}-\frac{\vec{\mu}}{2}\right|- \left|\vec{k}+\frac{\vec{\mu}}{2}\right|\right)ct\right\}\exp\left\{-i\vec{\mu}\cdot\vec{x}\right\}
$$
$$ \times 
\exp\left\{-i\phi_{m}(n-n')\right\}\widetilde{\Psi}_{n+1}\left(\vec{k}-\frac{\vec{\mu}}{2}\right)\widetilde{\Psi}_{n'+1}^{\ast} \left(\vec{k}+\frac{\vec{\mu}}{2}\right).
$$
It is obvious that
\begin{equation}\label{613}
  \rho_{W}=\Re \left( \Delta_{W} \right).
\end{equation}
\section{Concluding Remarks}
\label{sec7}
The main goal we set in this work was to construct the Weyl -- Wigner -- Moyal formalism for quantum mechanics of photon and to define the respective photon Wigner function within this formalism.

We have shown that such a program can be easily implemented within quantum mechanics for photon developed by I. Bia{\l}ynicki-Birula \cite{Bialynicki1994, Bialynicki1996, Bialynicki2006} and J. E. Sipe \cite{Sipe1995}, and with the use of the \textit{continuous-discrete} Wey -- Wigner -- Moyal formalism \cite{Przanowski2019, Przanowski2017} (see also the wide bibliography therein). The result reinforces our believe that one can successfully apply this formalism to other relativistic particles thus obtaining a promising approach to searching for the relativistic Wigner functions \cite{Bialynicki2014, Zavialov1999,Kowalski2016,Lev2001,Bialynicki1991,Shin1992,Best1993,Larkin2014}, although the problems with interpretation of the vector $\vec{x}$ are to be expected since for relativistic particles the operator $\widehat{\vec{x}}$ does not represent the position observable \cite{Hawton1999a,Davydov1976,Newton1949,Wightman1962,Pryce1948,Foldy1950,Kijowski1976,Hawton1999,Bliokh2017}. So the natural question is if one can construct the Weyl -- Wigner -- Moyal formalism for photon employing the position operator introduced by Margaret Hawton \cite{Hawton1999}.

 \section*{Acknowledgements}
 
 The work of F. J. T. was partially supported by SNI -- M\'{e}xico, COFAA -- IPN and by SIP -- IPN grant 20201186.

 \appendix

\section{Star product on the grid $\Gamma^3$}
\label{appendixA}
\setcounter{equation}{0}

The $*$ -- product on phase space $\Gamma$ \eqref{52} consists of two parts: a product on the  space ${\mathbb R}^6$ and a multiplication on the grid $\Gamma^3.$ Its general form is  
$$
(f*g)(\vec{p},\vec{x},\phi_{m},n)=
\sum_{k,l=0}^{2}\int_{\mathbb{R}^{6}}d^3\lambda d^3\mu\left(\mathcal{P}\left(\frac{\hbar \vec{\lambda}\cdot \vec{\mu}}{2}\right)\mathcal{K}\left(\frac{\pi k l}{3}\right)\right)^{-1}\exp\{i(\vec{\lambda} \cdot \vec{p} + \vec{\mu}\cdot \vec{x})\}
$$
\begin{equation}\label{a1}
\times
\exp\left\{ i\frac{2\pi}{3}(k \phi_m+ \phi_l n)\right\}(\tilde{\tilde{f}}\boxtimes\tilde{\tilde{g}})(\vec{\lambda}, \vec{\mu},k,l),
\end{equation}
where the $\boxtimes$ -- product does not depend on kernel ${\mathcal P}$ and ${\mathcal K}$.

Auxiliary functions $\tilde{\tilde{f}}(\vec{\lambda},\vec{\mu},k,l), \tilde{\tilde{g}}(\vec{\lambda},\vec{\mu},k,l)$ one calculates according to the rule
\[
\tilde{\tilde{f}}(\vec{\lambda}, \vec{\mu}, k,l)=  \frac{1}{(2 \pi)^6} \frac{1}{9} \,
\mathcal{P}\left(\frac{\hbar \vec{\lambda}\cdot \vec{\mu}}{2}\right)
{\mathcal K } \left( \frac{\pi k l}{3}\right) 
\]
\be
\label{a2}
\times
\int_{\mathbb{R}^{6}}d^3p d^3x
\sum_{m,n=0}^2
\exp\{-i(\vec{\lambda}\cdot \vec{p} + \vec{\mu}\cdot \vec{x})\}
\exp \left(-i \frac{2 \pi (k \phi_m+ \phi_l n)}{3} \right) f(\vec{p}, \vec{x}, \phi_m,n)
\ee
being a straightforward consequence of formula \eqref{534}.

The $\boxtimes$  -- multiplication on phase space ${\mathbb R}^6$ is defined as (compare \cite{Przanowski2019})
\[
\left( \tilde{\tilde{f}} \boxtimes \tilde{\tilde{g}} \right) (\vec{\lambda}, \vec{\mu}) = \int_{{\mathbb R}^{12}} d^3 \lambda' d^3 \mu' d^3 \lambda'' d^3 \mu''
\tilde{\tilde{f}}(\vec{\lambda}\,', \vec{\mu}\,')
\exp \left\{\frac{i \hbar}{2} (\vec{\lambda}\,' \cdot \vec{\mu} - \vec{\lambda} \cdot \vec{\mu}\,') \right\}
\]
\be
\label{a3}
\times
\delta(\vec{\lambda}\,' +\vec{\lambda}\,'' - \vec{\lambda}) \delta(\vec{\mu}\,' +\vec{\mu}\,'' - \vec{\mu})
\tilde{\tilde{g}}(\vec{\lambda}\,'', \vec{\mu}\,'').
\ee
 An explicit expression for the $\boxtimes $ -- product  \eqref{535} on the grid $\Gamma^3 $
 consists of the following set of terms
\[
\Big(\widetilde{ \widetilde {f} } \,\boxtimes \,\widetilde{ \widetilde {g} }\Big)(0,0)= 
\widetilde{ \widetilde {f} }(0,0)  \widetilde{ \widetilde {g} }(0,0) +
\widetilde{ \widetilde {f} }(0,2)  \widetilde{ \widetilde {g} }(0,1) +
\widetilde{ \widetilde {f} }(0,1)  \widetilde{ \widetilde {g} }(0,2) +
\widetilde{ \widetilde {f} }(2,0)  \widetilde{ \widetilde {g} }(1,0)
\]
\[ -
\widetilde{ \widetilde {f} }(2,2)  \widetilde{ \widetilde {g} }(1,1)+ 
\widetilde{ \widetilde {f} }(2,1)  \widetilde{ \widetilde {g} }(1,2) +
\widetilde{ \widetilde {f} }(1,0)  \widetilde{ \widetilde {g} }(2,0) +
\widetilde{ \widetilde {f} }(1,2)  \widetilde{ \widetilde {g} }(2,1)- 
\widetilde{ \widetilde {f} }(1,1)  \widetilde{ \widetilde {g} }(2,2), 
\]
\[
\Big(\widetilde{ \widetilde {f} } \,\boxtimes \,\widetilde{ \widetilde {g} }\Big)(0,1)= 
\widetilde{ \widetilde {f} }(0,1)  \widetilde{ \widetilde {g} }(0,0) +
\widetilde{ \widetilde {f} }(0,0)  \widetilde{ \widetilde {g} }(0,1) +
\widetilde{ \widetilde {f} }(0,2)  \widetilde{ \widetilde {g} }(0,2) +
\exp\left(\frac{2 i \pi}{3} \right)
\widetilde{ \widetilde {f} }(2,1)  \widetilde{ \widetilde {g} }(1,0)
\]
\[ - \exp\left(\frac{2 i \pi}{3} \right)
\widetilde{ \widetilde {f} }(2,0)  \widetilde{ \widetilde {g} }(1,1)+
\exp\left(\frac{2 i \pi}{3} \right) 
\widetilde{ \widetilde {f} }(2,2)  \widetilde{ \widetilde {g} }(1,2) +
\exp\left(\frac{ i \pi}{3} \right)
\widetilde{ \widetilde {f} }(1,1)  \widetilde{ \widetilde {g} }(2,0)
\]
\[
 -
\exp\left(\frac{ i \pi}{3} \right)
\widetilde{ \widetilde {f} }(1,0)  \widetilde{ \widetilde {g} }(2,1)- 
\exp\left(\frac{ i \pi}{3} \right)
\widetilde{ \widetilde {f} }(1,2)  \widetilde{ \widetilde {g} }(2,2), 
\]
\[
\Big(\widetilde{ \widetilde {f} } \,\boxtimes \,\widetilde{ \widetilde {g} }\Big)(0,2)= 
\widetilde{ \widetilde {f} }(0,2)  \widetilde{ \widetilde {g} }(0,0) +
\widetilde{ \widetilde {f} }(0,1)  \widetilde{ \widetilde {g} }(0,1) +
\widetilde{ \widetilde {f} }(0,0)  \widetilde{ \widetilde {g} }(0,2) +
\exp\left(-\frac{2 i \pi}{3} \right)
\widetilde{ \widetilde {f} }(2,2)  \widetilde{ \widetilde {g} }(1,0)
\]
\[ - \exp\left(-\frac{2 i \pi}{3} \right)
\widetilde{ \widetilde {f} }(2,1)  \widetilde{ \widetilde {g} }(1,1)+
\exp\left(-\frac{2 i \pi}{3} \right) 
\widetilde{ \widetilde {f} }(2,0)  \widetilde{ \widetilde {g} }(1,2) +
\exp\left(\frac{2 i \pi}{3} \right)
\widetilde{ \widetilde {f} }(1,2)  \widetilde{ \widetilde {g} }(2,0)
\]
\[
 -
\exp\left(\frac{2 i \pi}{3} \right)
\widetilde{ \widetilde {f} }(1,1)  \widetilde{ \widetilde {g} }(2,1)+
\exp\left(\frac{2 i \pi}{3} \right)
\widetilde{ \widetilde {f} }(1,0)  \widetilde{ \widetilde {g} }(2,2), 
\]

\[
\Big(\widetilde{ \widetilde {f} } \,\boxtimes \,\widetilde{ \widetilde {g} }\Big)(1,0)= 
\widetilde{ \widetilde {f} }(1,0)  \widetilde{ \widetilde {g} }(0,0) +
\exp\left(-\frac{2 i \pi}{3} \right)
\widetilde{ \widetilde {f} }(1,2)  \widetilde{ \widetilde {g} }(0,1) +
\exp\left(-\frac{ i \pi}{3} \right)
\widetilde{ \widetilde {f} }(1,1)  \widetilde{ \widetilde {g} }(0,2) 
\]
\[
+
\widetilde{ \widetilde {f} }(0,1)  \widetilde{ \widetilde {g} }(1,0)
 - \exp\left(-\frac{2 i \pi}{3} \right)
\widetilde{ \widetilde {f} }(0,2)  \widetilde{ \widetilde {g} }(1,1)-
\exp\left(-\frac{ i \pi}{3} \right) 
\widetilde{ \widetilde {f} }(0,1)  \widetilde{ \widetilde {g} }(1,2) +
\widetilde{ \widetilde {f} }(2,0)  \widetilde{ \widetilde {g} }(2,0)
\]
\[
 +
\exp\left(-\frac{2 i \pi}{3} \right)
\widetilde{ \widetilde {f} }(2,2)  \widetilde{ \widetilde {g} }(2,1)-
\exp\left(-\frac{ i \pi}{3} \right)
\widetilde{ \widetilde {f} }(2,1)  \widetilde{ \widetilde {g} }(2,2), 
\]

\[
\Big(\widetilde{ \widetilde {f} } \,\boxtimes \,\widetilde{ \widetilde {g} }\Big)(1,1)= 
\widetilde{ \widetilde {f} }(1,1)  \widetilde{ \widetilde {g} }(0,0) +
\exp\left(\frac{ i \pi}{3} \right)
\widetilde{ \widetilde {f} }(1,0)  \widetilde{ \widetilde {g} }(0,1) +
\exp\left(-\frac{ i \pi}{3} \right)
\widetilde{ \widetilde {f} }(1,2)  \widetilde{ \widetilde {g} }(0,2) 
\]
\[
+\exp\left(-\frac{ i \pi}{3} \right)
\widetilde{ \widetilde {f} }(0,1)  \widetilde{ \widetilde {g} }(1,0)
+
\widetilde{ \widetilde {f} }(0,0)  \widetilde{ \widetilde {g} }(1,1)-
\exp\left(-\frac{2 i \pi}{3} \right) 
\widetilde{ \widetilde {f} }(0,2)  \widetilde{ \widetilde {g} }(1,2) 
\]
\[
+
\exp\left(\frac{ i \pi}{3} \right)
\widetilde{ \widetilde {f} }(2,1)  \widetilde{ \widetilde {g} }(2,0)
 -
\exp\left(\frac{2 i \pi}{3} \right)
\widetilde{ \widetilde {f} }(2,0)  \widetilde{ \widetilde {g} }(2,1)-
\widetilde{ \widetilde {f} }(2,1)  \widetilde{ \widetilde {g} }(2,2), 
\]

\[
\Big(\widetilde{ \widetilde {f} } \,\boxtimes \,\widetilde{ \widetilde {g} }\Big)(1,2)= 
\widetilde{ \widetilde {f} }(1,2)  \widetilde{ \widetilde {g} }(0,0) +
\exp\left(\frac{ i \pi}{3} \right)
\widetilde{ \widetilde {f} }(1,1)  \widetilde{ \widetilde {g} }(0,1) +
\exp\left(\frac{2 i \pi}{3} \right)
\widetilde{ \widetilde {f} }(1,0)  \widetilde{ \widetilde {g} }(0,2) 
\]
\[
+\exp\left(-\frac{2 i \pi}{3} \right)
\widetilde{ \widetilde {f} }(0,2)  \widetilde{ \widetilde {g} }(1,0)
+ \exp\left(-\frac{ i \pi}{3} \right)
\widetilde{ \widetilde {f} }(0,1)  \widetilde{ \widetilde {g} }(1,1)+
\widetilde{ \widetilde {f} }(0,0)  \widetilde{ \widetilde {g} }(1,2) 
\]
\[
+
\exp\left(\frac{2 i \pi}{3} \right)
\widetilde{ \widetilde {f} }(2,2)  \widetilde{ \widetilde {g} }(2,0)
 +
\widetilde{ \widetilde {f} }(2,1)  \widetilde{ \widetilde {g} }(2,1)+
\exp\left(-\frac{2 i \pi}{3} \right)
\widetilde{ \widetilde {f} }(2,0)  \widetilde{ \widetilde {g} }(2,2), 
\]

\[
\Big(\widetilde{ \widetilde {f} } \,\boxtimes \,\widetilde{ \widetilde {g} }\Big)(2,0)= 
\widetilde{ \widetilde {f} }(2,0)  \widetilde{ \widetilde {g} }(0,0) +
\exp\left(\frac{2 i \pi}{3} \right)
\widetilde{ \widetilde {f} }(2,2)  \widetilde{ \widetilde {g} }(0,1) +
\exp\left(-\frac{2 i \pi}{3} \right)
\widetilde{ \widetilde {f} }(2,1)  \widetilde{ \widetilde {g} }(0,2) 
\]
\[
+
\widetilde{ \widetilde {f} }(1,0)  \widetilde{ \widetilde {g} }(1,0)
- \exp\left(-\frac{2 i \pi}{3} \right)
\widetilde{ \widetilde {f} }(1,2)  \widetilde{ \widetilde {g} }(1,1)-
\exp\left(-\frac{2 i \pi}{3} \right)
\widetilde{ \widetilde {f} }(1,1)  \widetilde{ \widetilde {g} }(1,2) 
\]
\[
+
\widetilde{ \widetilde {f} }(0,0)  \widetilde{ \widetilde {g} }(2,0)
 + \exp\left(\frac{2 i \pi}{3} \right)
\widetilde{ \widetilde {f} }(0,2)  \widetilde{ \widetilde {g} }(2,1)+
\exp\left(-\frac{2 i \pi}{3} \right)
\widetilde{ \widetilde {f} }(0,1)  \widetilde{ \widetilde {g} }(2,2), 
\]

\[
\Big(\widetilde{ \widetilde {f} } \,\boxtimes \,\widetilde{ \widetilde {g} }\Big)(2,1)= 
\widetilde{ \widetilde {f} }(2,1)  \widetilde{ \widetilde {g} }(0,0) +
\exp\left(\frac{2 i \pi}{3} \right)
\widetilde{ \widetilde {f} }(2,0)  \widetilde{ \widetilde {g} }(0,1) +
\exp\left(-\frac{2 i \pi}{3} \right)
\widetilde{ \widetilde {f} }(2,1)  \widetilde{ \widetilde {g} }(0,2) 
\]
\[
+ \exp\left(-\frac{ i \pi}{3} \right)
\widetilde{ \widetilde {f} }(1,1)  \widetilde{ \widetilde {g} }(1,0)
- \exp\left(\frac{2 i \pi}{3} \right)
\widetilde{ \widetilde {f} }(1,0)  \widetilde{ \widetilde {g} }(1,1)+
\widetilde{ \widetilde {f} }(1,2)  \widetilde{ \widetilde {g} }(1,2) 
\]
\[
+ \exp\left(-\frac{2 i \pi}{3} \right)
\widetilde{ \widetilde {f} }(0,1)  \widetilde{ \widetilde {g} }(2,0)
 + 
\widetilde{ \widetilde {f} }(0,0)  \widetilde{ \widetilde {g} }(2,1)+
\exp\left(\frac{2 i \pi}{3} \right)
\widetilde{ \widetilde {f} }(0,2)  \widetilde{ \widetilde {g} }(2,2), 
\]

\[
\Big(\widetilde{ \widetilde {f} } \,\boxtimes \,\widetilde{ \widetilde {g} }\Big)(2,2)= 
\widetilde{ \widetilde {f} }(2,2)  \widetilde{ \widetilde {g} }(0,0) +
\exp\left(\frac{2 i \pi}{3} \right)
\widetilde{ \widetilde {f} }(2,1)  \widetilde{ \widetilde {g} }(0,1) +
\exp\left(-\frac{2 i \pi}{3} \right)
\widetilde{ \widetilde {f} }(2,0)  \widetilde{ \widetilde {g} }(0,2) 
\]
\[
+ \exp\left(-\frac{2 i \pi}{3} \right)
\widetilde{ \widetilde {f} }(1,2)  \widetilde{ \widetilde {g} }(1,0)
+
\widetilde{ \widetilde {f} }(1,1)  \widetilde{ \widetilde {g} }(1,1)+
\exp\left(\frac{2 i \pi}{3} \right)
\widetilde{ \widetilde {f} }(1,0)  \widetilde{ \widetilde {g} }(1,2) 
\]
\[
+ \exp\left(\frac{2 i \pi}{3} \right)
\widetilde{ \widetilde {f} }(0,2)  \widetilde{ \widetilde {g} }(2,0)
 + \exp\left(-\frac{2 i \pi}{3} \right)
\widetilde{ \widetilde {f} }(0,1)  \widetilde{ \widetilde {g} }(2,1)+
\widetilde{ \widetilde {f} }(0,0)  \widetilde{ \widetilde {g} }(2,2). 
\]

\end{document}